\documentclass[structabstract]{aa}  
\usepackage{float}
\usepackage{graphicx}
\usepackage{longtable}
\usepackage{txfonts}
\begin{document}
\title{An updated stellar census of the Quintuplet cluster
\thanks{Based on observations made at the European Southern Observatory, 
Paranal, Chile under programme ESO 093.D-0306} 
}
\author{J.~S.~Clark\inst{1}
\and M.~E.~Lohr\inst{1}
\and L.~R.~Patrick\inst{2,3,4}
\and F.~Najarro\inst{5}
\and H.~Dong\inst{6}
\and D.~F.~Figer\inst{7} 
}
\institute{
$^1$School of Physical Science, The Open 
University, Walton Hall, Milton Keynes, MK7 6AA, United Kingdom\\
$^2$Instituto de Astrof\'{\i}sica de Canarias, E-38205 La Laguna, Tenerife, Spain\\ 
$^3$Universidad de La Laguna, Dpto Astrof\'{i}sica, E-38206 La Laguna, Tenerife,
Spain\\
$^4$Universidad de La Laguna, Dpto Astrofı́sica, E-38206 La Laguna, Tenerife,
Spain\\
$^5$Departamento de Astrof\'{\i}sica, Centro de Astrobiolog\'{\i}a, 
(CSIC-INTA), Ctra. Torrej\'on a Ajalvir, km 4,  28850 Torrej\'on de Ardoz, 
Madrid, Spain\\
$^6$Instituto de Astrof\'{i}sica de Andaluc\'{i}a (CSIC), Glorieta de la
Astronom\'{a} S/N, E-18008 Granada, Spain\\
$^7$Center for Detectors, Rochester Institute of Technology, 74 Lomb Memorial Drive, Rochester, NY 14623, USA 
0000-0002-4206-733X}

    \abstract{The Quintuplet  is one of the most massive young clusters in the Galaxy. 
 As a consequence it offers the prospect of constraining stellar  formation and  evolution in extreme
 environments. However, current observations suggest that it comprises a remarkably diverse stellar population that is difficult 
to reconcile with an instantaneous formation event.}
 {To better understand the nature of the cluster we aim  to improve observational constraints on the constituent stars.}{In 
 order to accomplish this goal we present  HST/NICMOS+WFC3 photometry and VLT/SINFONI+KMOS spectroscopy for $\sim100$ and 71  cluster members, respectively. }
 {The Quintuplet appears far more homogeneous than previously expected.  All 
 supergiants are classified as either  O7-8 Ia or O9-B0 Ia, with only one object of earlier (O5 I-III) spectral type. These 
 stars form a smooth morphological sequence with a cohort of seven early-B hypergiants and six luminous blue variables and 
 WN9-11h stars, which comprise the richest population of such stars of any stellar aggregate known. In parallel, we 
 identify a smaller population of late-O hypergiants and spectroscopically similar WN8-9ha stars. No further H-free Wolf-Rayet (WR)
 stars were identified, resulting in a 13:1 ratio  for WC/WN stars. A subset of the O9-B0 supergiants are unexpectedly 
 faint, suggesting they are both less massive and older than the greater cluster population.}
 {Due to an uncertain extinction law, it is not possible to quantitatively determine a cluster age via 
 isochrone fitting. Nevertheless, we find an impressive coincidence between the properties of  cluster members preceding the 
 H-free WR phase and the evolutionary predictions for a single, non-rotating $60M_{\odot}$ star, implying  an  age of 
 $\sim3.0-3.6$Myr. Neither the  late O-hypergiants nor the low luminosity supergiants are predicted by such a 
 path; we suggest that the former either result from  rapid rotators or are the products of binary driven mass-stripping, while 
 the latter may be interlopers. The H-free WRs must evolve from stars with an initial mass in excess of $60M_{\odot}$ but it 
 appears difficult to reconcile their observational properties with theoretical expectations. This is important since one would 
 expect the most massive stars within the Quintuplet to be undergoing core-collapse/SNe at this time; since the WRs represent 
 an evolutionary phase directly  preceding this event,their physical properties are crucial to understanding both this process 
 and the nature of the resultant relativistic remnant. As such, the Quintuplet provides unique observational constraints on the 
evolution and death of the most massive stars forming  in the local, high metallicity Universe.

}


\maketitle

\section{Introduction}

The formation and subsequent lifecycle of very massive stars is one of the outstanding problems of stellar physics, 
impacting on  fields as diverse as the re-ionisation in the early Universe, the chemical enrichment of galaxies, and  the 
production of both electromagnetic and gravitational wave transients. Recent observations suggest that the formation of 
such stars is, in many cases, hierarchical, with binaries and higher-order multiples forming in stellar clusters or  
associations, which in turn are located within cluster complexes. 
Such  star-forming structures are characteristic of  many external (starburst) galaxies; unfortunately,  whilst it is 
possible to  resolve and study individual  clusters, one cannot  identify their  consituent stars. As a 
consequence one must rely on  observations of  local clusters and  associations in order to benchmark the population and 
spectral synthesis codes used to infer the properties of such aggregates at cosmological distances.

Found within $\sim200$pc of the Galactic Centre and  characterised by predominantly molecular rather than atomic gas, the
central molecular zone (CMZ) is of particular interest. The presence of 
numerous compact H\,{\sc ii} regions within the Sgr B2 molecular cloud (e.g. De Pree et al. \cite{depree}), three young 
($<10$Myr) massive ($\gtrsim10^4M_{\odot}$) clusters (YMCs) - the Arches, Quintuplet, and circumnuclear Galactic Centre 
cluster (e.g. Figer et 
al. \cite{figer99b}, Paumard et al. \cite{paumard06}) - as well as a population of apparently isolated massive stars (e.g. 
Mauerhan et al. \cite{mauerhan10a}, \cite{mauerhan10c}, Dong et al. \cite{dong11}, \cite{dong12}, \cite{dong15}) 
points to dramatic recent and ongoing star formation within the CMZ. However, despite containing some $\sim80$\% of the 
 dense molecular gas within the Galaxy (Morris \& Serabyn \cite{MS}) and  with individual clouds apparently of 
sufficient mass and density to form clusters such as the Arches (e.g.  G0.253+0.016; Longmore et al. \cite{longmore}),  
the rate of star formation (per unit mass) within the CMZ appears depressed by at least an order of magnitude 
 in comparison to that inferred for the Galactic disc (Barnes et al. \cite{barnes}).

The reason(s) for this discrepancy are unclear  (cf. Barnes et al. \cite{barnes}), but given this 
observation and the extreme conditions present within the CMZ\footnote{The average density, pressure, temperature, and  
(turbulent) velocity dispersion of the gaseous material as well as the strength of the magnetic fields threading clumps, 
the frequency of clump/clump  collisions,  the  intensity of the ambient radiation 
field and the cosmic ray ionisation rate are all significantly higher than found within the Galactic disc (cf. Barnes et 
al. \cite{barnes}, Kauffmann \cite{kauf}).} one might ask whether clusters and their constituent stars form in the same 
manner and have the same physical properties as their counterparts in more quiescent regions of the Galactic disc. This is  
an important question since circumnuclear starbursts are present in many external galaxies and the conditions within the 
CMZ are thought to be similar to those present within high redshift starburst galaxies (cf. Barnes et al. \cite{barnes}, but 
see also Kauffmann \cite{kauf}).
In order to address these questions we have initiated a reappraisal of the properties of both the isolated and clustered 
stellar populations present within the CMZ. Previous publications (Clark et al. \cite{clark18}, Lohr et al. \cite{lohr18}) have 
concentrated on the Arches, and in this paper we turn our attention to the Quintuplet.

 Initial near-IR surveys of the Galactic centre revealed the Quintuplet as a heavily blended source. However, in
1990  three groups successfully resolved
the Quintuplet into multiple stellar sources (Glass et al. \cite{glass}, Nagata et al. \cite{nagata90} and
Okuda et al. \cite{okuda}), with the latter two noting the extreme luminosity and cool spectral energy distribution of five 
members of the putative cluster. Subsequent spectroscopic confirmation of the presence of
massive stars within the Quintuplet was provided by  a number of different authors  (e.g. Moneti et al. \cite{moneti94}, 
Geballe et al. \cite{geballe94}, Figer et al. \cite{figer95},\cite{figer96}, Cotera et al. \cite{cotera}), with
extensive spectroscopic surveys published by Figer et al. (\cite{figer99a}; henceforth Fi99a) and Liermann et al. 
(\cite{liermann09}; 
henceforth Li09). In conjunction with photometric data, these observations suggested that the Quintuplet was  both older 
than the  Arches - e.g. $3\pm0.5$My (Liermann et al. \cite{liermann12}) to $4\pm1$Myr (Fi99a) versus 
$\sim2-3$Myr (Clark et al. \cite{clark18}) -  and substantially less dense, although integrated masses were expected to be 
broadly comparable (Figer et al. \cite{figer99b}).

These results, particularly the divergent cluster densities,  prompt the question of whether the  two clusters 
formed via different mechanisms  or if secular evolution and/or an   interaction with an outside agent such as the tidal 
field of the Galactic centre has led to this situation. Indeed, the large range  of spectral types and luminosity classes 
apparently exhibited by Quintuplet members (e.g. Li09, Liermann et al. \cite{liermann12}) raises the  possibility that the 
 cluster is not co-eval and  instead results from  multiple episodes of star formation or the merger of sub-clusters of 
differing ages. Schneider et al.  (\cite{schneider14}) investigated this issue and were able to demonstrate that, despite 
these findings, the 
Quintuplet  could be co-eval, but at the cost of inferring an (interacting) binary fraction of $\sim60$\% and an age 
somewhat larger than commonly assumed ($4.8\pm1.1$Myr).

The possibility of a significant  role for  binary interaction in  the evolution of the constituent stars of the 
Quintuplet heralds a further observational opportunity. Current spectral analysis suggest that it hosts a rich and diverse 
population of post-main  sequence (MS)  objects (Fi99a, Li09) including comparatively rare objects such as luminous blue 
variables  (LBVs) and OB hypergiants. Despite the brevity of these phases, the extreme mass loss rates that characterise 
them are expected to profoundly influence the evolution of massive stars  by stripping away the H-rich 
mantle to reveal the nuclear 
processed,  chemically enriched core. If the  cluster formation history may be adequately quantified, and the properties 
of the putative binary population constrained, the Quintuplet  will serve as an excellent testbed for stellar evolution 
theory for both single and binary channels,
providing constraints that   complement those derived from both   younger and older clusters such as the Arches and 
Westerlund 1, respectively (Clark et al. \cite{clark05a}, \cite{clark18}, Negueruela et al. \cite{negueruela}). 

In order to address these issues,this paper presents a (re)analysis of Hubble Space Telescope (HST)/NICMOS+WFC3 
near-IR photometry and Very Large Telescope (VLT)/SINFONI+KMOS spectroscopy for the Quintuplet. In Sect. 2 we describe  
data acquisition and the 
reduction  techniques employed before presenting the resultant datasets. Sect. 3 sets out the spectral classification of 
cluster members, with the implications of these for global cluster properties highlighted in Sect. 4. A discussion of issues arising from these findings is presented in Sect. 5, before we draw our final conclusions in Sect. 6.

\section{Data acquisition, reduction and analysis}

\begin{figure*}[!ht]
\includegraphics[width=15.5cm,angle=0]{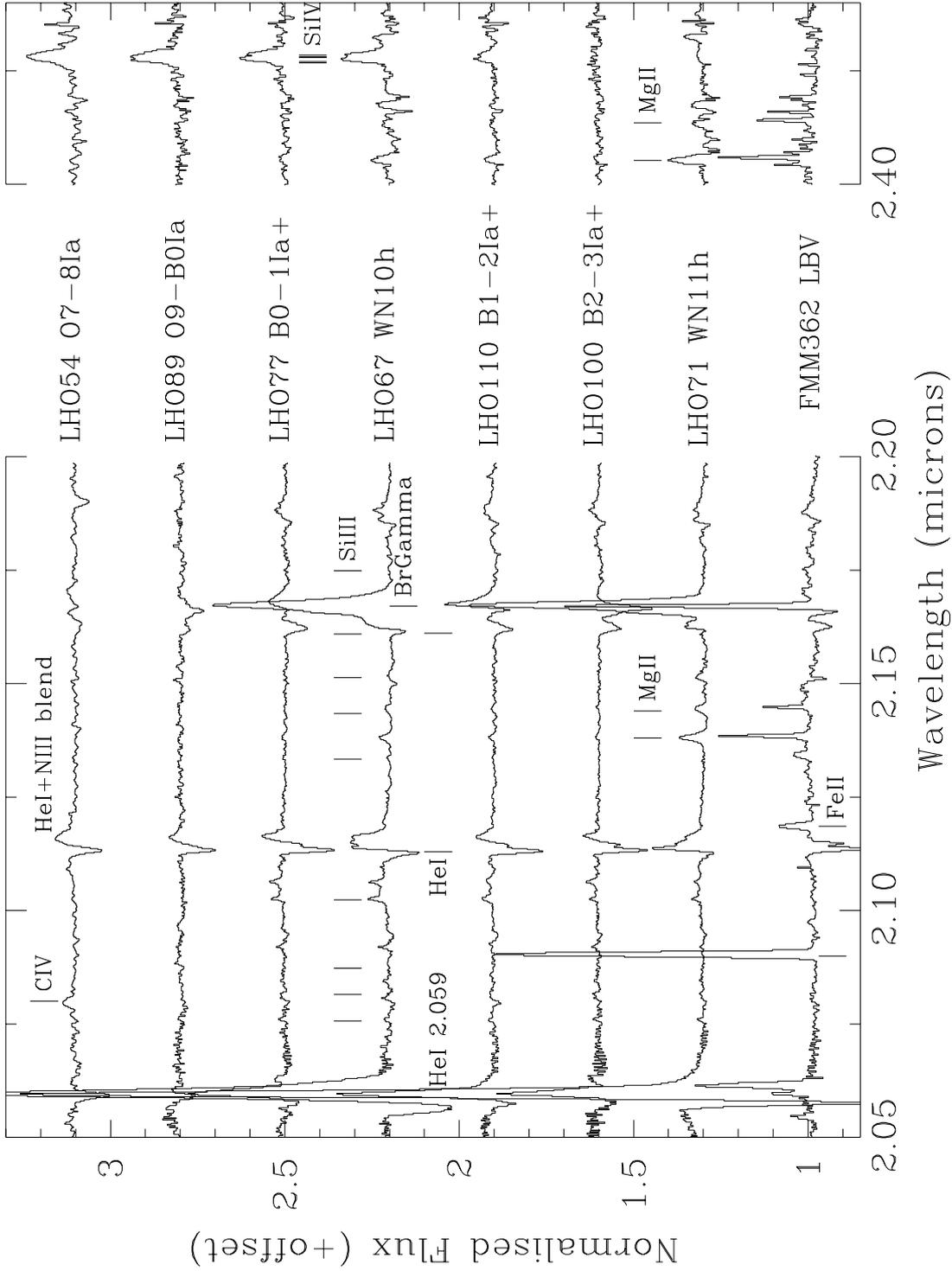}
\caption{Montage of spectra of selected O supergiant, B hypergiant, WNVLh, and LBV stars to illustrate the evolution in strengths of the Si\,{\sc iv}, Si\,{\sc iii}, and Mg\,{\sc ii} lines  as a function of decreasing stellar temperature. In order to render these transitions visible, the  He\,{\sc i}2.058$\mu$m and Br$\gamma$ profiles of certain stars overlap; these are reproduced in Figs. 6 and 7.
Note the appearance of pronounced emission in Fe\,{\sc ii} in the coolest star, the LBV qF362. See Sect. 3.1 for further details.}
\end{figure*}

\subsection{Spectroscopy}

Data from VLT/KMOS and VLT/SINFONI were reduced following the methodology described below. The resultant spectra are presented in Figs. 1-9, A.1, and A.2.

\subsubsection{VLT/KMOS data}
The VLT-KMOS (Sharples et al. \cite{sharples}) data for this paper
were obtained under  ESO programme 093.D-0306 (PI: Clark),
with observations made between 2014 August 02-13. 
The spectral resolution of the observations is a function of rotator angles and the integral field units (IFUs) used (cf. Patrick et al. \cite{patrick15}),
 varying between ${\Delta}{\lambda}/{\lambda}\sim3895 - 4600$. Each observing block consisted of 12$\times$30\,s exposures in an ABA
observing pattern, where the first observation of each field used the more
rigorous 24-arm telluric standard star approach and all subsequent
observations of the same field used the standard 3-arm telluric approach.
The standard stars used for these observations were HIP\,84846 (A0V),
HIP\,91137 (A0V), and HIP\,3820 (B8V).

Science and standard star observations were calibrated, reconstructed and
combined using the KMOS/esorex pipeline (Davies et al. \cite{kmos}).
In the K-band the sky emission lines present a significant problem and to
apply an accurate and consistent sky subtraction we used the KMOS/esorex
pipeline with the \textit{sky\_tweak} option.
The wavelength solution for each extracted science and telluric standard
star spectrum was checked and improved upon using an iterative
cross-correlation approach, where a high resolution spectrum of the Earth's
telluric absorption was used as a reference\footnote{Retrieved from
http://eso.org/sci/facilities/paranal/decommissioned/isaac/
tools/spectroscopic\_standards.html}.

As the majority of the useful diagnostic lines for these targets lie in a region of the K-band that is 
highly contaminated by telluric absorption, we
implemented a rigorous telluric correction routine, adapted from Patrick et
al. (\cite{patrick15}, \cite{patrick17}). Specifically, to create a 
telluric spectrum free of stellar absorption features, the
Br~$\gamma$ absorption line present in the standard star was
modelled with a double Lorentzian profile, via an iterative approach tailored to each KMOS IFU,
in order to simultaneously fit the
wings and centre of the profile in order to allow the most complete removal of it.

Once the Br~$\gamma$ correction is applied to an acceptable degree, the
telluric standard star spectra are all continuum normalised by identifying
multiple continuum points, where telluric contamination is minimised, across
the entire spectral range (1.934 -- 2.460\,$\mu$m).
A linear interpolation to these data creates a continuum spectrum,
which is then used to divide each telluric spectrum.
The science observations are continuum normalised in a similar fashion
using, however, an entirely different set of continuum points tailored
individually to the spectral appearance of the star in question.

The final telluric correction is made when the science and telluric spectra
are continuum normalised.
In order to best match the telluric spectrum to the science, a scaling
function is applied to the former, which is defined by using multiple wavelength 
regions where the telluric absorption is most severe.
An appropriate scaling factor is identified for each region individually by
iteratively comparing the science and telluric spectra.
The scaling function is then defined for the entire wavelength range by
means of a linear interpolation between the multiple indiviudal  scaling factors.
After the scaling function has been applied to the telluric spectrum, the
science spectrum is divided by the resulting telluric spectrum.
The telluric-corrected science spectrum is then continuum normalised in the
fashion outlined above, using a linear interpolation to several
continuum points which are defined  on the basis of  
spectral appearance.

In total this programme resulted in spectra of 19 cluster members, of which ten were in common with the VLT/SINFONI dataset of Li09
and nine unique to the VLT/KMOS observations. 

\subsubsection{VLT/SINFONI data}

\begin{figure*}[!ht]
\includegraphics[width=12cm,angle=0]{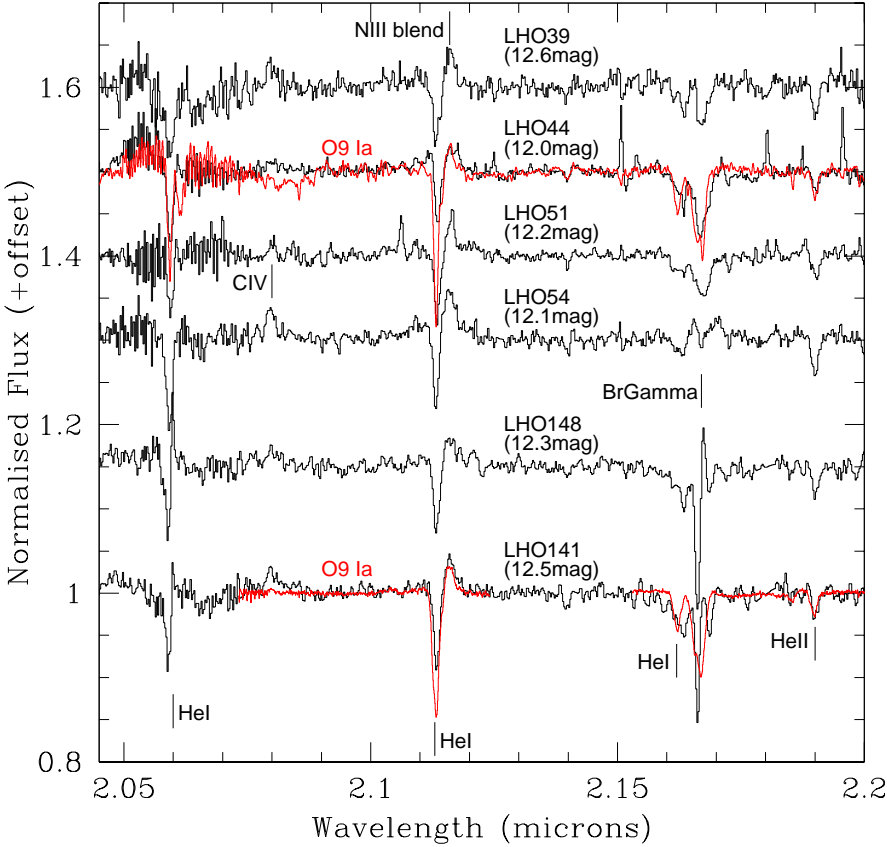}
\caption{Examples of cluster supergiants (black) which appear earlier than O9-B0 Ia template spectra (red; Hanson et al. \cite{hanson05}) by virtue of
ubiquitous and pronounced He\,{\sc ii} 2.189$\mu$m absorption and C\,{\sc iv} 2.079$\mu$m emission. Even when the strength of  He\,{\sc ii} 2.189$\mu$m absorption is comparable to the O9 Ia template (e.g. LHO44), the combination of 
 C\,{\sc iv} 2.079$\mu$m emission and/or much weaker He\,{\sc i} 2.112$\mu$m absorption clearly marks them out as of earlier spectral type. We therefore adopt a classification of O7-8 Ia, noting that the broad Br$\gamma$ emission in LHO54 leads to a classification of  O7-8 Ia$^{(+)}$ (Sects. 3.2 and 3.3).  HST/NICMOS F205W magnitudes are given in parentheses (Table A.1).}
\end{figure*}

\begin{figure*}[!ht]
\includegraphics[width=12cm,angle=0]{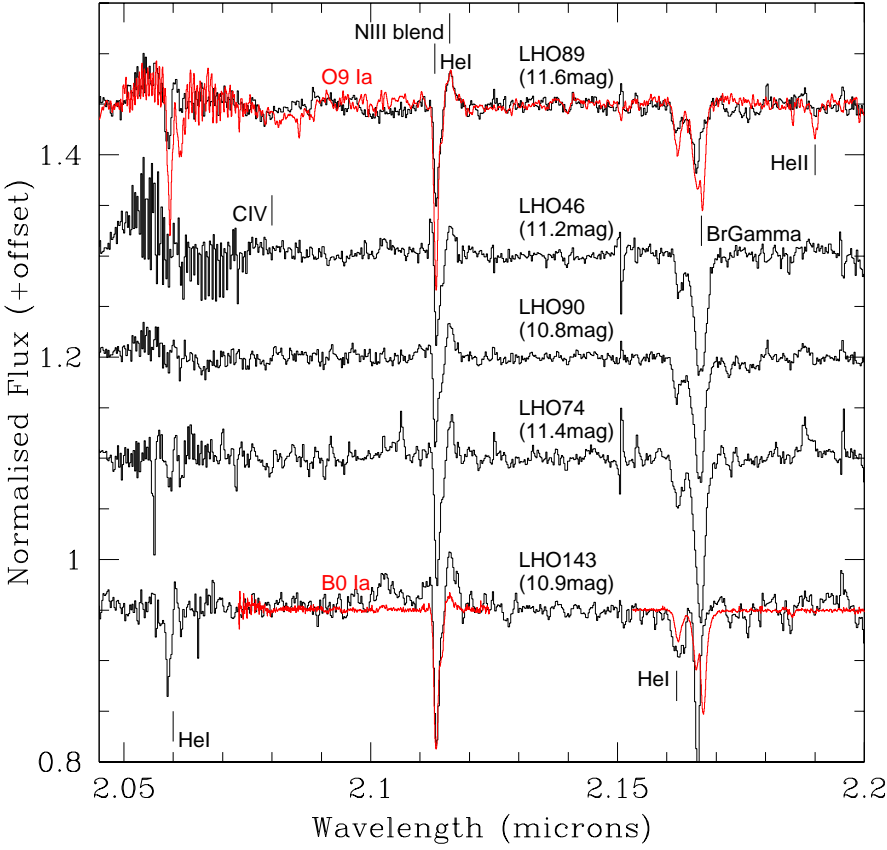}
\caption{Examples of cluster supergiants (black) which we may classify as O9-B0 Ia due to their close correspondence
to appropriate template spectra (red) from Hanson et al. (\cite{hanson05}). HST/NICMOS F205W magnitudes are given 
in parentheses (Table A.1).}
\end{figure*}

\begin{figure}[!ht]
\includegraphics[width=7cm,angle=-90]{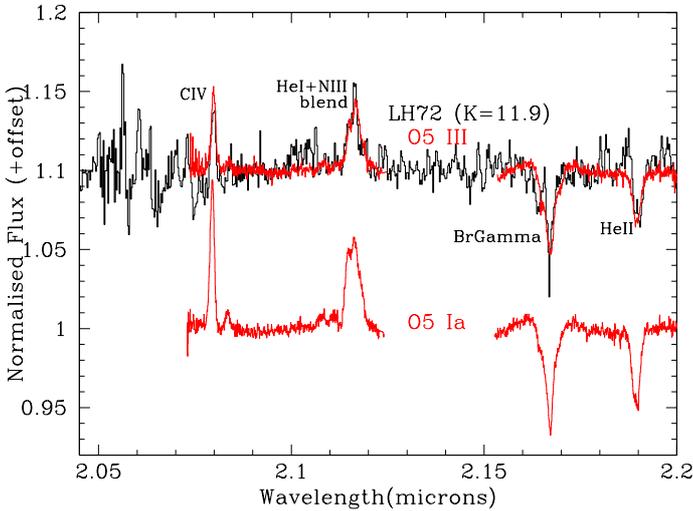}
\caption{The spectrum of LHO72 (black), the earliest O star detected within the Quintuplet and O5 Ia and O5 III template spectra (red) for comparison. The HST/NICMOS F205W magnitude is given 
in parenthesis (Table A.1).}
\end{figure}

\begin{figure*}[!ht]
\includegraphics[width=12cm,angle=0]{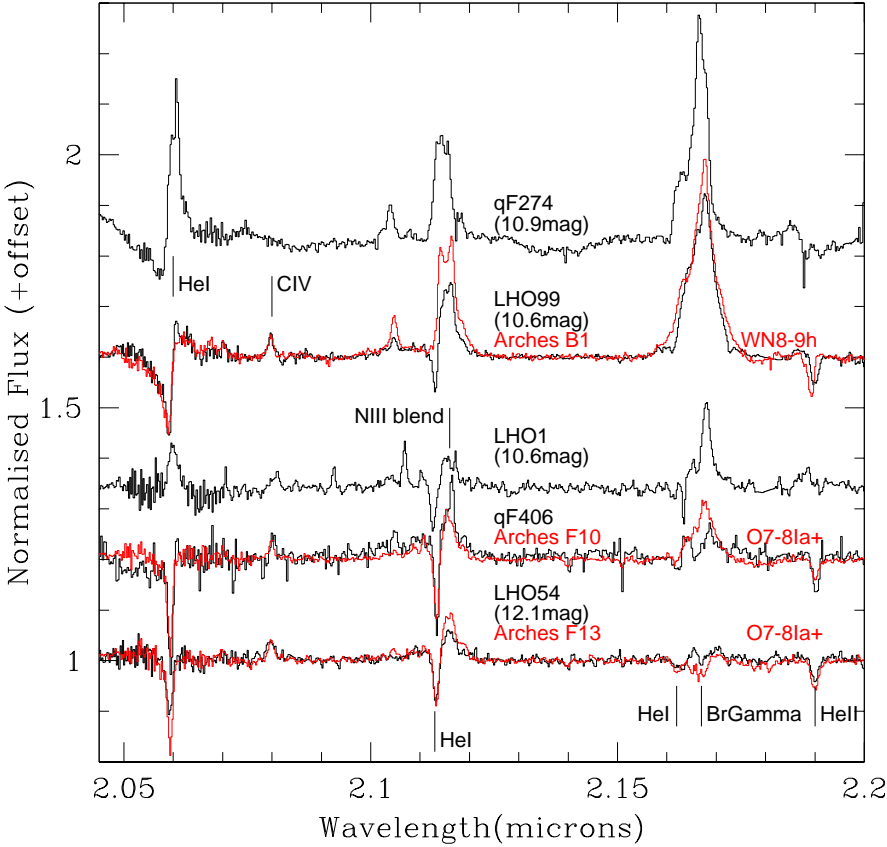}
\caption{Spectra of late O hypergiant and WN8-9h cluster members (black) in comparison to spectra of stars within
the Arches cluster (red; Clark et al. \cite{clark18}). Where available HST/NICMOS F205W magnitudes are given 
in parentheses (Table A.1).}
\end{figure*}

\begin{figure*}[!ht]
\includegraphics[width=16.5cm,angle=0]{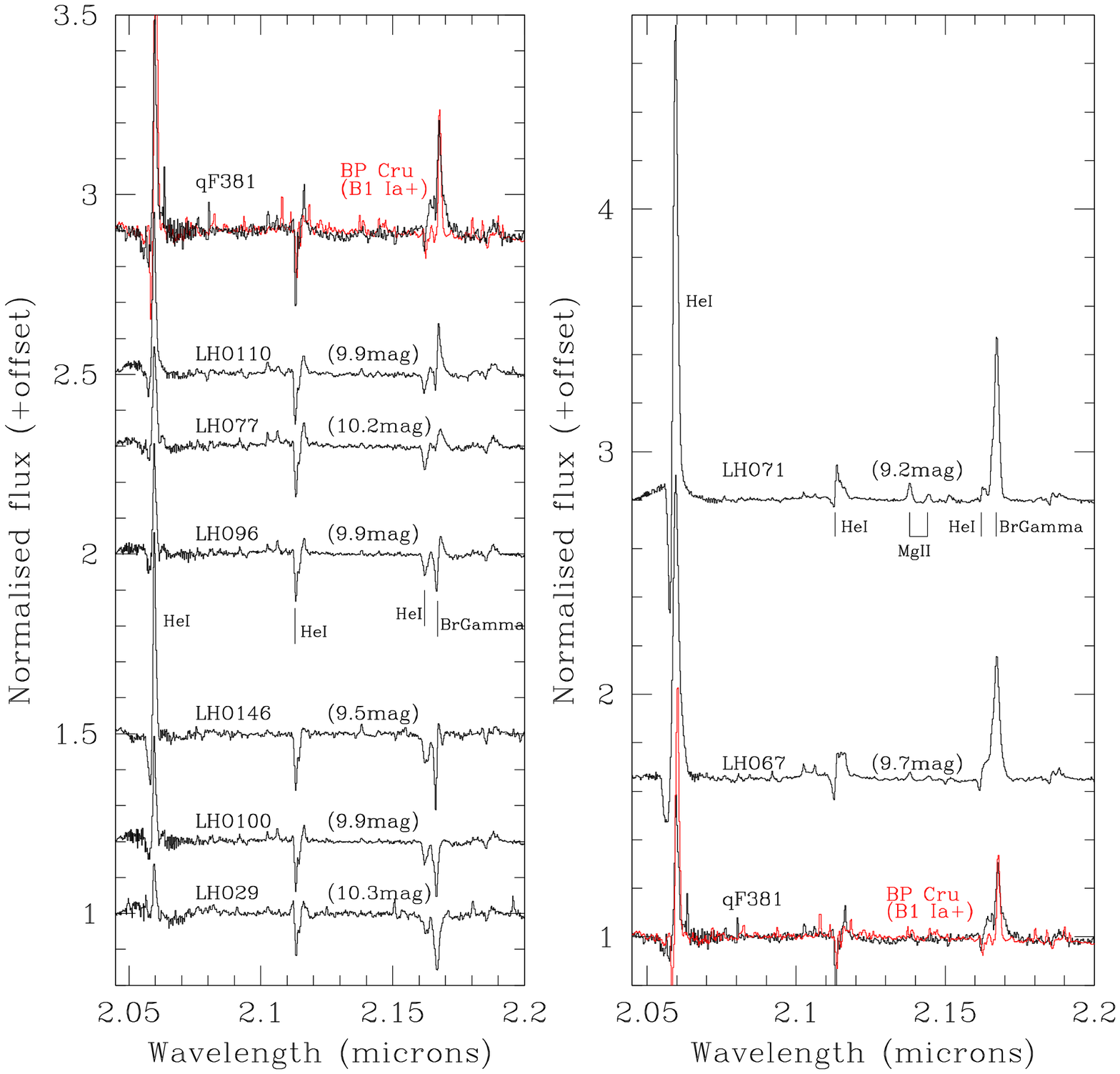}
\caption{{\em Left panel}: montage of spectra of early-B hypergiants within the Quintuplet. The spectrum of the
B1 Ia$^+$ hypergiant BP Cru is provided for comparison (red; spectrum from Waisberg et al. \cite{waisberg}). {\em Right panel}: comparison of the B0-1 Ia$^+$ star qF381 to the cluster WN10h star LHO67 and the  WN11h star LHO71. Given the scale only prominent emission lines are shown; weaker transitions are indicated in Fig. 1.
Where available HST/NICMOS F205W magnitudes are given 
in parentheses (Table A.1).
}   
\end{figure*}

Spectroscopic observations were made in service mode\footnote{ESO proposal 077.D-0281(A)} of the central region of the Quintuplet cluster using the SINFONI integral field spectrograph on the ESO/VLT (Eisenhauer et al. 2003; Bonnet et al. 2004) between 18 May and 1 July 2006 (see Li09 for a full description of the instrument set-up, field placements and times of observation).  The raw science and telluric frames and associated calibration files were downloaded from the ESO archive and reduced with the latest version of the ESO SINFONI pipeline running under Reflex, which performed flat-fielding and optical distortion corrections, wavelength calibration and improved sky background subtraction, before constructing co-added data cubes for each observation.

As noted by Li09, the sky fields chosen contained obvious stars, contaminating the spectra of the sky-subtracted science fields in corresponding regions and in extreme cases producing negative stellar spectra in certain pixels of the science cubes.  However, the majority of each sky-subtracted science cube was not significantly affected and was very well cleaned of OH sky emission lines by the new SINFONI pipeline software.  Therefore, a flattened sky image was used to generate a bad pixel mask for each sky-subtracted science frame, and, with a version of the custom IDL code described in Clark et al. (\cite{clark18}), multiple uncontaminated spectral pixels were manually selected for each target of interest, and optimally combined to produce a single spectrum per object.  This approach produced cleaner final spectra than the alternative methods of subtracting a ``monochromatic median'' of a starless region of each science cube (as carried out by Li09), or replacing the star-affected pixels of the sky images by median spectra derived from the whole sky cube prior to subtraction from the science cubes.  A few fainter targets of interest were fully coincident with stars in the corresponding sky fields, and so were not extracted here, since no significant improvement in their spectra was to be expected relative to those published in Li09.  Objects with obvious CO bandhead absorption (classified by Li09 as K--M giants or supergiants) were not generally extracted, on the presumption that they were not cluster members and that further re-reduction would not reveal additional spectral features.
Likewise we refrained from extracting spectra of obviously blended and/or particularly faint stars (foreshadowing Sect. 2.2 the faintest star for which a spectrum was extracted had $m_{\rm F205W}\sim14.7$).

Science spectra had small corrections made to wavelength calibration, and telluric absorption line removal was carried out, as described fully in Sect. 2.1.1 above and  Clark et al. (\cite{clark18}). Continuum normalisation again followed the methodology outlined above and in that work. Barycentric corrections were then applied, and, in the few cases where two observations had been made of a single object\footnote{Where field placements overlapped - objects LHO67, -69, -71, -75, and -110.} the two final spectra were averaged together.  Bad lines (present at around 2.028 and 2.180$\mu$m in almost all spectra) were manually removed; a few other unphysical features resulting from nebular emission in the sky fields or from imperfect telluric correction were manually removed by interpolation, after careful comparison of science spectra to telluric or sky spectra to ensure the feature in question was spurious and not astrophysical in origin. 

However two fields were subject to nebular contamination in both He\,{\sc i} 2.059$\mu$m and Br$\gamma$ which we were unable to remove in a robust and objective manner from the spectra of a number of stars\footnote{LHO5, -9, -143, -145, -152, -154, -157, and -159.}. Foreshadowing Sect. 3 both transitions are important classification diagnostics. Additional diagnostics in regions free of nebular contamination suggested early spectral-types  for both LHO5 and -143; unfortunately the low S/N of the remaining objects precluded meaningful analysis, save that they do not appear to be WR stars due to their lack of strong broad emission lines. As a consequence we do not discuss these stars further. We suspect residual contamination in the Br$\gamma$ emission profile of LHO16 and the absorption feature of LHO149, although these are retained since it proved possible to classify them.

This resulted in a much reduced dataset of spectra for 54 stars compared to the 160 presented in Li09 (noting that 62 of these were of stars of spectral-type K or M and hence were not re-extracted).  However the  methodology employed here greatly improved the S/N of the resultant spectra, allowing us to access weaker emission (e.g. C\,{\sc iv} 2.079$\mu$m)  and absorption (e.g. He\,{\sc ii} 2.189$\mu$m) features that are critical to spectral classification of massive stars (Sect.3).

\subsection{Photometry}

Currently, the only photometric data available for indivdual stars in the literature is derived from ground-based
observations (e.g. Fi99a, Li09). Given the limitations of such data in terms of crowding and atmospheric 
transmission we  compiled HST/NICMOS+WFC3 observations of potential members of the Quintuplet.
 These are of both greater sensitivity and angular  
 resolution than previous studies and critically  also permit the construction of spectral energy distributions and subsequent 
 de-reddening of individual stars, as well as direct comparison to members of the Arches  and isolated massive stars across the 
CMZ for which HST data are also available.

  HST/NICMOS observations of the Quintuplet in the F110W, F160W, and F205W filters were made in 1997 (GO 7362; PI Figer)  
and are described and  presented in Figer et al. (\cite{figer99b}). We refer the reader to that work for details of reduction 
and analysis. Unfortunately photometric data associated with  individual stars was not presented in that work, and so 
here we provide such a breakdown by cluster member (Table A.1). Complimentary and hitherto unpublished  HST/WFC3 photometry in the F127M, F139M, and F153M filters was obtained in  2010-2012 under programmes GO-11671, 12318, and 12667 (PI, Andrea Ghez); a detailed description of data acquisition, reduction and analysis, including error determination, can be found in  Dong et al. (\cite{dong17}). 

An initial  target list was derived from the catalogues of stellar sources given in Fi99a and Li09, with the exclusion of the K-M stars which are likely interlopers. This contained a number of sources from Li09 that are found within the core  of the cluster and for which we were unable to re-extract and process unblended spectra; these were retained to inform future observational programmes.
In order to match sources from both HST programmes  the coordinate system derived from the WFC3 dataset was applied to the NICMOS image. A star  was considered detected in the NICMOS dataset if it lay within 2.5pixel of its expected position as determined from the WFC3
observations. This resulted in a total of 108 sources, for which photometry is presented in Table A.1. Of these, four WFC3 sources lack NICMOS
photometry, while the NICMOS counterpart is uncertain for a further five objects due to the presence of more than one object within the search 
radius. These are flagged within the comments of Table A.1, as are eight further stars for which spectral classifications are available but 
photometry is unavailable due to blending or their falling outside the fields-of-view of the observations.

Given the very reddened nature of the stars within the Quintuplet, the differing filter passbands and the possibility of blending we refrain from comparing the resultant dataset to published ground based observations (Fi99a, Li09). Likewise we highlight that the NICMOS and WFC3 data were obtained over a decade apart and that a number of stars are either known  to be variable (e.g. the LBVs; Glass et al. \cite{glass99}, Mauerhan et al. \cite{mauerhan10a}) or might be anticipated to be based on other examples (e.g. the WNLh and WCL cohorts); thus caution should be applied in constructing and interpreting spectral energy distributions from these data.

\section{Spectral classification}

In combination the VLT/SINFONI+KMOS datasets yield spectra of 63 unique stars. To these we may add historical spectra of a further eight stars. Six of these originate in Fi99a, while the seventh and eighth, of the candidate late-O supergiant qF344 and the  LBV G0.120-0.048 (= [DWC2011] 92)\footnote{Which,  given its proximity and similarity to both the Pistol star and qF362 we consider likely to have originated in the Quintuplet (cf. Mauerhan et al. (\cite{mauerhan10b})}, are from Mauerhan et al. (\cite{mauerhan10a}, \cite{mauerhan10b}).  This results in a spectroscopic census of 71 putative massive cluster members. Unlike our Arches survey (Clark et al. \cite{clark18}) it is not anticipated that these observations will be sensitive  enough to reliably reach the main sequence and giant cluster populations. 
Moreover, as a consequence of their low spectral resolution, S/N and  amplitude of spectral features present, we treat classifications derived from spectra presented by Fi99a as provisional\footnote{Stars previously identified as late-O/early-B supergiants have subsequently been classified as 
late-type stars (qF269), LBVs (qF362), WCLs (qF250), and mid-late O and early-B hypergiants (qF257, qF270S, qF278, qF381, and qF406).}.

\subsection{Methodology}

\begin{figure*}[!ht]
\includegraphics[width=16.5cm,angle=0]{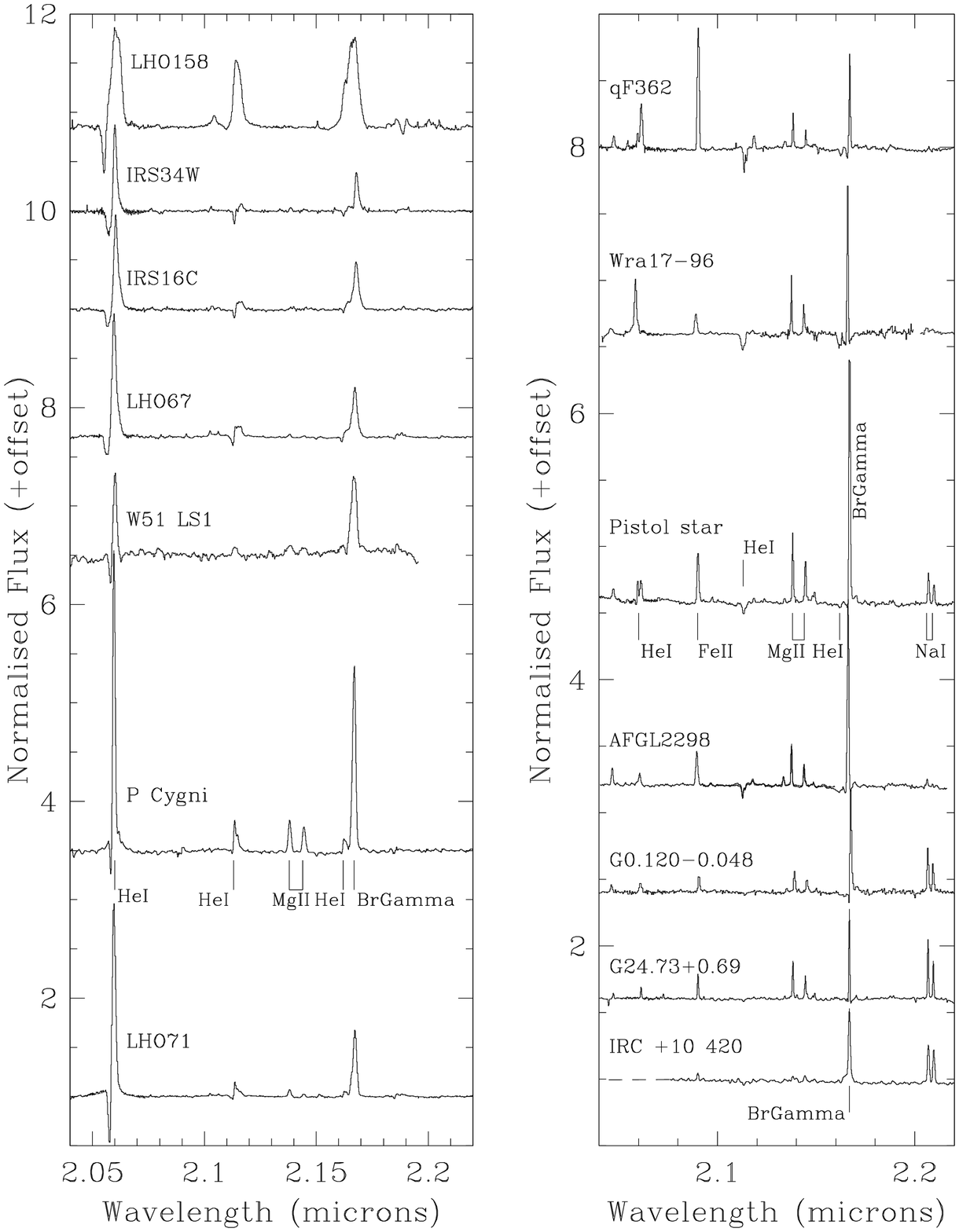}
\caption{Montage of spectra of cluster LBVs (qF362, the Pistol star and G0.120-0.048) in comparison to field LBVs and WNLh stars. Spectra of the latter from 
Martins et al. (\cite{martins07}; GC IRS 34W and 16C), Clark et al. (\cite{clark09a}; W51 LS1), Najarro (\cite{paco01}; P Cygni), Egan et al. (\cite{egan}; Wra17-96), Clark et al. (\cite{clark09b}; AFGL 2298), this work (footnote 11; G24.73+0.69), and Yamamuro et al. (\cite{yamamuro}; IRC +10 420). Given the scale only prominent emission lines are shown; weaker transitions are indicated in Fig. 1. Given that Mauerhan et al. (\cite{mauerhan10b}) demonstrate that all three LBVs are variable we refrain from presenting HST data for these stars. }
\end{figure*}

As with the Arches, uncertainty in the correct extinction law to apply and likely differential reddening across the Quintuplet means that we rely solely on spectroscopic data for classification purposes. A detailed reprise of 
spectral diagnostics in the K-band window was provided in Clark et al. (\cite{clark18}); we do not reproduce this here for reasons of space. However, given the more diverse nature of the stellar population of the Quintuplet we briefly describe relevant publications utilised for classification and salient details of the procedure if not contained within that work. 

Hanson et al. (\cite{hanson05}) provide classification criteria for OB supergiants although, foreshadowing Sect. 3.2, there is an unfortunate absence of template spectra of spectral type $\sim$O6.5-8. We employ data from the Arches supplemented by the low spectral resolution and S/N data from Hanson et al. (\cite{hanson96}) to infer trends in spectral morphologies for supergiants in this range. 
No one publication presents classification criteria for OB hypergiants; for early- to mid-O hypergiants we utilise Hanson et al. (\cite{hanson05}) and Clark et al. (\cite{clark18}), while spectra and analysis of  late-O/early-B hypergiants are provided by Clark et al. 
(\cite{clark12}, \cite{clark14b}) and Waisberg et al. (\cite{waisberg}). Representative spectra of LBVs in both hot- and cool-phases are provided by Morris et al. (\cite{morris}), Clark et al. (\cite{clark03},  \cite{clark05b}, \cite{clark09a}, \cite{clark11})
  and Oksala et al. (\cite{oksala}). Wolf-Rayet classification criteria are provided by Figer et al. (\cite{figer97}), Crowther et al. (\cite{crowther06}), Crowther \& Walborn (\cite{crowther11}) and Rosslowe \& Crowther (\cite{rosslowe}).

However there are a number of additional diagnostics not discussed in these works that are newly accessible to us via our 
greatly 
improved telluric correction and sky subtraction. The most prominant of these is a strong emission feature at $\sim2.43\mu$m. Non-LTE model atmosphere simulations utilising the CMFGEN code (Hillier \& Miller \cite{hillier98},\cite{hillier99}) suggest the identity of this feature is sensitive to temperature, with contributions from, respectively, the n=$10{\rightarrow}9$  lines  of O\,{\sc iv}, N\,{\sc iv}, C\,{\sc iv}, and finally Si\,{\sc iv} as one transitions to cooler temperatures, with the latter  dominating for the stars 
considered here. Given this one would also expect the emission feature to  demonstrate an explicit abundance dependence  as the products of CNO burning are revealed at the stellar surface.

Following the discussion in Clark et al. (\cite{clark18}) for WNLh stars hotter  than 
$\sim30$kK one would expect Si\,{\sc iv} to dominate the emission feature. N\,{\sc iv} starts to contribute at $\sim32$kK,
 equalling the strength of Si\,{\sc iv} at $36.5$kK.
 Subject to depletion C\,{\sc iv} might be expected to provide a minor contribution; since the stars considered here are almost certainly cooler than  $\sim45$kK one would not expect a contribution from  O\,{\sc iv}. Similarly for supergiants of spectral type mid-O and later  one would expect this feature to result from a combination of Si\,{\sc iv}, C\,{\sc iv}, and N\,{\sc iv}, with Si\,{\sc iv} increasingly dominant for cooler stars.

 Simulations indicate that the Si\,{\sc iv} $\sim2.427\mu$m feature drops out around spectral types $\sim$B1.5-2 ($\sim18-20$kK); high S/N spectra of stars in this temperature range  which drive dense winds  reveals the presence of  additional weak Si\,{\sc iii} 2.0367, 2.0746, 2.0805, 2.0863, 2.1013, and  2.1323$\mu$m emission lines\footnote{Additional Si\,{\sc iii} emission and absorption features are present within the K-band spectral window; here we simply reproduce the subset that are present in LHO67 and are not blended with other stronger lines.}. Emission also develops in the  Mg\,{\sc ii} 2.1369, 2.1432, 2.4042, and 2.4125$\mu$m lines, which  become particularly prominent for spectral type B3 and later 
($\lesssim15-16$kK). Finally, for comparatively  cool stars ($\lesssim10$kK), absorption in the He\,{\sc i} 2.112$\mu$m line disappears and the Na\,{\sc i} 2.206/9$\mu$m transitions may be seen in emission (cf. Clark et al. 
\cite{clark11}). 

The spectra of Quintuplet members presented in Fig. 1 show the expected trends in line strengths for  
Si\,{\sc iv} $\sim2.427\mu$m,   Mg\,{\sc ii} 2.138, 2.144, and 2.4125$\mu$m and the multiple 
Si\,{\sc iii} transitions, while the weakening of He\,{\sc i} 2.112$\mu$m absorption and development 
 of Na\,{\sc i} 2.206/9$\mu$m emission at yet cooler temperatures is apparent in the montage of LBV spectra in Fig. 7. The classifications of individual spectra in these figures is discussed further below.

Supergiant B[e] (sgB[e]) stars complete the menagerie of hot and  massive evolved stars relevant to the Quintuplet.
Such stars are of considerable interest as they are thought to be masive binaries either undergoing (e.g. Wd1-9; Clark et al. \cite{clark13}, Fenech et al. \cite{fenech}), or to have recently exited from, rapid (case-A) mass transfer (e.g. Kastner et al. \cite{kastner}, Clark et al. \cite{clark13a}).
 Spectra of sgB[e] stars are presented by Clark et al. (\cite{clark99}), Oksala et al. (\cite{oksala}) and Liermann et al. (\cite{liermann14}); superficially these resemble those of LBVs, being dominated by emission from hydrogen, helium and low excitation metallic species. However a large number also present CO bandhead in emission and, by definition, all show a pronounced near-IR excess 
due to hot circumstellar dust; no stars within the Quintuplet fulfil all these criteria.

Finally  we turn to stars of later (F-M) spectral type.  Liermann et al. (\cite{liermann12}) identify a {\em possible} population of 
low mass ($<9M_{\odot}$) red supergiants via pronounced CO bandhead absorption (Davies et al. \cite{davies}) but, given their discrepant radial velocities, they conclude that these are likely interlopers. The most luminous - LHO7 (=MGM 5-7) - has a mass
of $\sim15-20M_{\odot}$, significantly less than we infer for  {\em bona fide} cluster members (Sect. 4) suggesting it too is unlikely to be a cluster member. More massive ($\geq30-40M_{\odot}$) RSGs - such as those within 
Westerlund 1 (Borgmann et al. \cite{borgmann}, Westerlund \cite{westerlund}) - 
would support near-IR luminosities that would rival or exceed the Quintuplet-proper members and consequently we can conclude that such stars are not present. Likewise, comparison of the spectra of cluster members to those of  yellow hypergiants (Yamamuro et al. \cite{yamamuro}, Clark et al. \cite{clark14a}) show these too are absent.

\subsection{OB supergiants}

Accurate classification of the supergiant cohort is vital given the implications it has for both the age and the co-evality of the Quintuplet. Consequently, one of the more surprising results of Li09 is that the hot supergiants they identify appear to span an unexpectedly wide range of spectral types, from as early as O3 (e.g. LHO1,  -17, -65, and -128) to B3 (LHO32 and  -37)  although many are classified within a more restricted range of spectral types 
($\sim$O7-B1). 

Fortunately, the higher S/N afforded by our combined spectroscopic dataset affords greater confidence in the presence, or otherwise, of weak, temperature dependent classification criteria such as C\,{\sc iv} 2.079$\mu$m, He\,{\sc i} 2.112$\mu$m, and He\,{\sc ii} 2.189$\mu$m,  as well as permitting a robust interpretation of the blended He\,{\sc i} 2.162$\mu$m+Br$\gamma$ profile, which is sensitive to stellar luminosity as well as temperature.  As can be seen from Figs. 2 and 3, the majority of supergiants can be divided into two distinct morphological groups on the basis of these diagnostics.

The first consists of 18 supergiants (with $m_{\rm F205W}\sim12-13.1$) that are characterised by 
 C\,{\sc iv} 2.079$\mu$m emission and/or  He\,{\sc ii} 2.189$\mu$m absorption (including the transitional super-/hypergiant LHO54; Figs. 2 and A.1 and Table 1). The presence of the former and the weakness of the  temperature-sensitive He\,{\sc i} 2.112$\mu$m absorption feature in comparison to the template spectra of Hanson et al. (\cite{hanson05}) suggests a spectral-type earlier than O9 Ia. Conversely the presence of He\,{\sc i} 2.112$\mu$m absorption and the comparative weakness of C\,{\sc iv} 2.079$\mu$m emission suggests that these stars are cooler than the O4-6 Ia supergiants that characterise the Arches cluster (Martins et al. \cite{martins08}, Clark et al. \cite{clark18}). Mindful of the lack of appropriate template spectra, we consequently adopt an intermediate classification of O7-8 Ia for these stars. An additional argument for this assignment is the depth of Br$\gamma$ absorption in many of these stars, being markedly more pronounced than the He\,{\sc ii} 2.189$\mu$m line.  

Emission is present  in the He\,{\sc i} 2.162$\mu$m+Br$\gamma$ blend of a number of these stars, albeit of differing nature - e.g. the broad emission in LHO54 (see also Sect. 3.3.1) versus the narrow emission components in the red wing of Br$\gamma$ in LHO141 and -148. Additionally a degree of infilling appears present in LHO31, -73, -118, and -144. He\,{\sc i} 2.059$\mu$m is seen in absorption, with narrow emission in the red wing again seen in stars such as LHO141 and -148; mirroring the behaviour of Br$\gamma$ and arguing for the presence of stronger winds in these objects.

The second cohort comprises nine stars which  closely match the O9-B0 Ia template spectra from Hanson et al. (\cite{hanson05}; Figs. 3 and A.2 and Table 1). These uniformly lack C\,{\sc iv} 2.079$\mu$m emission and the He\,{\sc ii} 2.189$\mu$m photospheric profile is very weak or absent. Pronounced absorption is seen in the He\,{\sc i}  2.112$\mu$m line and the He\,{\sc i} 2.162$\mu$m+Br$\gamma$ blend, although emission in the red wing of the Br$\gamma$ profile may be present in some stars (e.g. LHO89). The 
He\,{\sc i} 2.059$\mu$m profile is complex, with apparent infilling or narrow emission components present in a number of stars (e.g. LHO90 and -143). Five of these objects (Fig. 3) appear more luminous than the O7-8 Ia cohort, with $m_{\rm F205W}\sim10.9-11.6$; unexpectedly  the remaining four (Fig. A.2) appear systematically fainter with $m_{\rm F205W}\sim13.0-13.8$, despite having comparable near-IR colours (hence apparently excluding differential reddening as an explanation). 

Additionally, six rather faint objects ($m_{\rm F205W}\sim13.3-14.7$) have low S/N spectra that are dominated by a pronounced Br$\gamma$ photospheric profile, with He\,{\sc i} 2.059$\mu$m and/or 2.112$\mu$m absorption also present in some cases. The Br$\gamma$ profiles appear systematically too deep for them to be early-mid O-type stars (cf. Martins et al. \cite{martins08}, Clark et al. \cite{clark18}), while they  also lack the pronounced emission in the  2.11$\mu$m blend that is ubiquitous in such stars. Subject to the low S/N, pronounced He\,{\sc ii} 2.189$\mu$m emission also appears to be absent. As a consequence, while we assign a generic OB star classification, we suspect these are unlikely to have a  spectral type earlier than late-O.

 Li09 suggest that a number of stars potentially have rather early classifications (e.g. LHO88 and -128; O3-4 IIIf   and O3-5 I, respectively). Our results do not support this hypothesis (cf. Table 1), with  LHO72 appearing to be of the earliest spectral type of our sample (Fig. 4): the comparable strength of emission in C\,{\sc iv} 2.079$\mu$m and  the He\,{\sc i}+N\,{\sc iii} $\sim2.11\mu$m blend, when combined with the lack of He\,{\sc i} 2.112$\mu$m absorption and the strong He\,{\sc ii} 2.189$\mu$m photospheric profile, suggesting an O5 classification. Differentiating between luminosity class 
I-III is difficult based solely on spectral morphology, although with $m_{\rm F205W}\sim12.9$ LHO72 is broadly comparable to the fainter subset of the  O7-8 supergiants discussed above

Finally we turn to the six stars for which only archival spectra are available. Comparing the spectrum of qF344 from Mauerhan et al. (\cite{mauerhan10a}) to our dataset suggests a {\em tentative} O7-8 Ia clasification on the basis of the strong He\,{\sc ii} absorption although, with $m_{\rm F205W}\sim11.5$, it is somewhat brighter than other examples (possibly suggesting a marginally later classification).
  Likely due to their low S/N and resolution, the spectra of four of the remaining five stars\footnote{qF157 ($<$B0 Ia, $m_{\rm F205W}\sim11.9$), qF276 (B1-3 Ia, $m_{\rm F205W}\sim11.0$),  qF307 (B1-3 Ia, $m_{\rm F205W}\sim11.9$),  qF311 (B1-3 Ia), and   qF358 (B1-3 Ia, $m_{\rm F205W}\sim10.5$)} appear essentially featureless; the exception being  qF307, which appears to 
have He\,{\sc i} 2.059$\mu$m in emission. The classifications suggested by Fi99a are broadly  consistent with the wider supergiant cohort and their near-IR magnitudes. A possible exception is qF307; the combination of its brightness and spectral morphology suggesting a potential early-B hypergiant classification (cf. qF257, -278, and -381; footnote 7 and Sect. 3.3).

\subsection{Hypergiants}

\begin{figure*}[!ht]
\includegraphics[width=13cm,angle=-0]{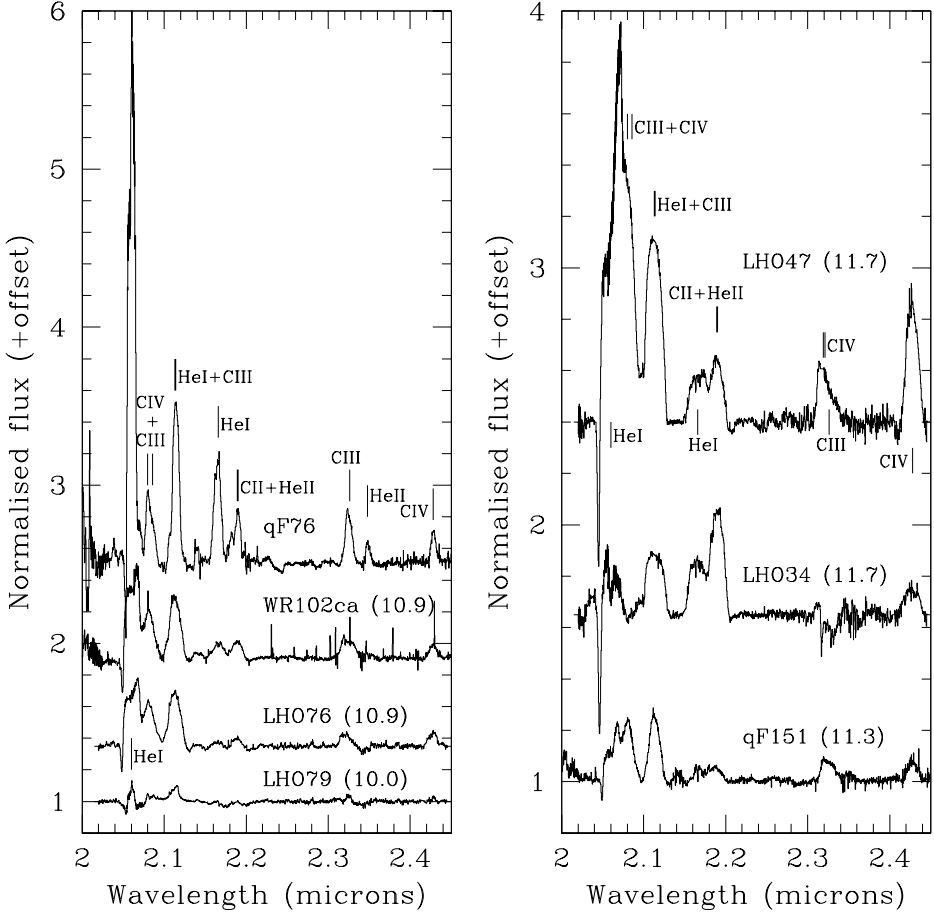}
\caption{Montage of cluster WC spectra. {\em Left panel:} WC9 stars and {\em right panel:} WC8-9 and WC8 stars.
Where available HST/NICMOS F205W magnitudes are given in parentheses (Table A.1), although we note that dusty WCL are known to be photometrically variable.
}
\end{figure*}

\begin{figure*}[h]
\includegraphics[width=11cm,angle=-0]{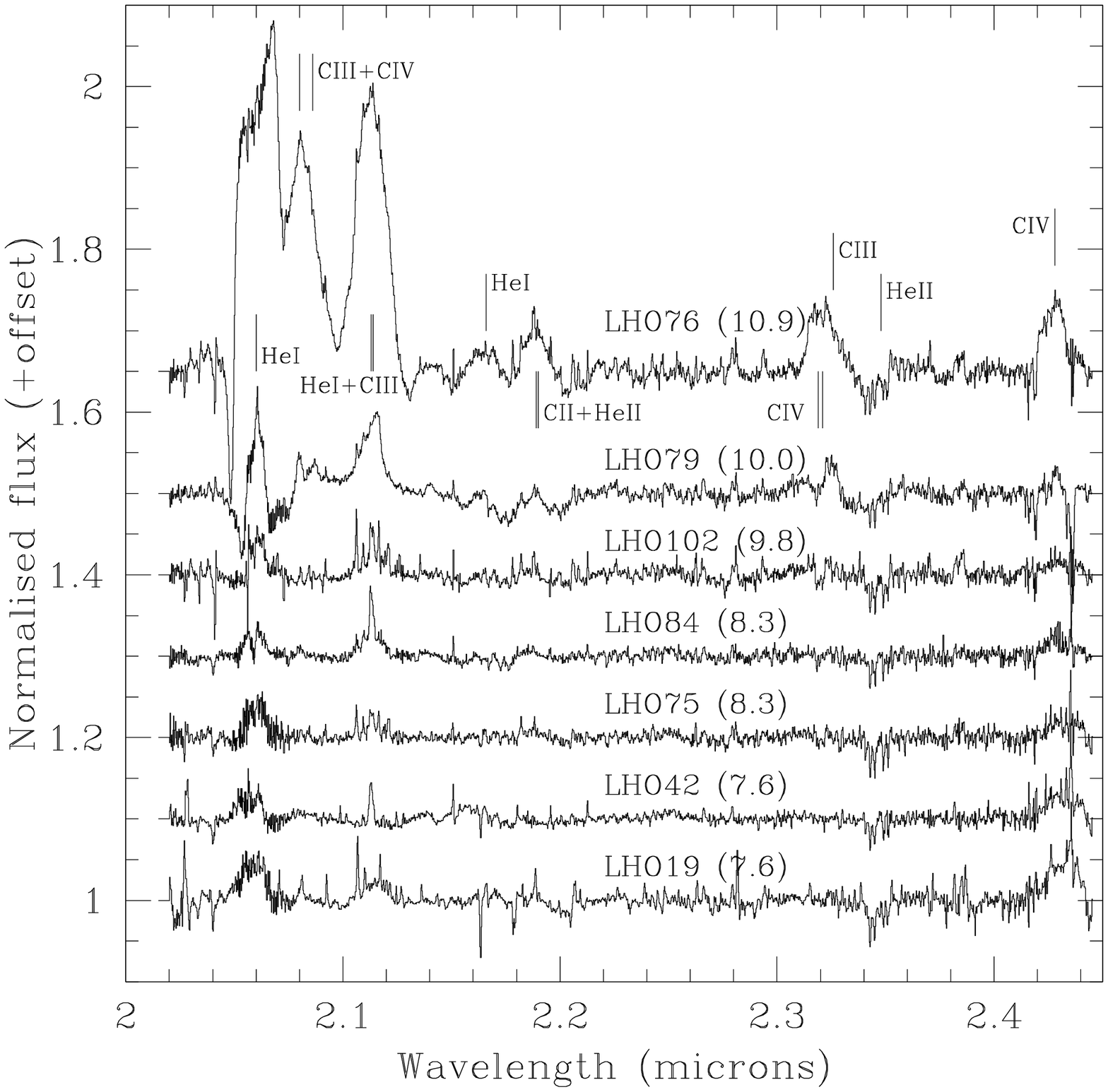}
\caption{Montage of WC stars with weaker emission features due to dilution by emission from hot dust. LHO76 and -79 are shown again for ease of comparison. Note that sharp emission features between e.g. 2.1$\mu$m and 2.11$\mu$m and visible in LHO19,-42 -75, -84, and -102 are residual sky lines; we find no evidence for the putative secondaries in any spectra.  
 HST/NICMOS F205W magnitudes are given in parentheses (Table A.1), although we note that dusty WCL are known to be photometrically variable. Finally for ease of comparison the five Quintuplet proper members are LHO19, -42, -75, -84, and -102 (=MGM 5-3, -2, -1, -4, and -9 respectively). }
\end{figure*}

Our data suggest the presence of two distinct cohorts of hypergiants within the Quintuplet, distinguished by their temperatures; we discuss these separately below.

\subsubsection{O hypergiants}
As described in Sect. 3.2, a subset of stars with spectral features consistent with spectral type O7-8
show broad emission  in Br$\gamma$, indicative of a more developed stellar wind than generally seen in the supergiant population. 
Of these both LHO54 and qF406 bear close resemblance to the hypergiants within the Arches cluster (Fig. 5), with the former apparently intermediate between qF406 and the Quintuplet supergiants. LHO1 appears more extreme, with emission also 
present  in He\,{\sc i} 2.059$\mu$m. As a consequence we adopt an O7-8 Ia$^+$  classification for LHO1 and qF406 and 
O7-8 Ia$^{(+)}$ for LHO54 (Table 1).

\subsubsection{B hypergiants}

Li09 identifies a subset  of five stars - LHO77, -96, -100, -110, and -146 -  defined by the presence of strong, narrow emission in  He\,{\sc i} 2.059$\mu$m and absorption in the He\,{\sc i} 2.112$\mu$m line.  Following our (re-)reduction of the combined datasets we may add LHO29 and qF381 to this grouping (Fig. 6). Li09 define these as O6-8 Ife stars but comparison of these spectra to those of similar spectral type in both the Quintuplet and Arches (Figs. 1, 2, and 5; Clark et al. \cite{clark18}) does not support such a classification. Specifically they lack C\,{\sc iv} 2.079$\mu$m  emission and  He\,{\sc ii} 2.189$\mu$m absorption, while the He\,{\sc i} 2.059 $\mu$m emission profile is anomalous. Such a combination suggests substantially cooler temperatures, moreso with the identification of weak Si\,{\sc iii} features in a number of stars in conjunction with the weakening and disappearance of Si\,{\sc iv} 2.427$\mu$m emission. Specifically qF381, LHO77, and LHO110 show the strongest Si\,{\sc iv} emission, with further tentative identifications in LHO29, -96,  and -146 and an absence in LHO100, which we interpret as a progression to cooler temperatures (Fig. 1).

Comparison to template spectra indicates that emission in He\,{\sc i} 2.059$\mu$m and Br$\gamma$ is atypical for supergiants but consistent with early-B hypergiants (Clark et al. \cite{clark12}, \cite{clark14b}, Waisberg et al. \cite{waisberg}). Consequently we classify these stars as B0-1 Ia$^+$ (qF381, LHO77), B1-2 Ia$^+$ (LHO29, -96, -110, and -146) and B2-3 Ia$^+$ (LHO100). Liermann et al. (\cite{liermann10}) have already drawn comparison between LHO110 and the WN9h stars within the cluster; our new spectrum of qF381, with more pronounced  He\,{\sc i} 2.059$\mu$m and 
Br$\gamma$ emission,  provides an even more compelling illustration of this morphological progression (Fig. 6).  At the other extreme, one may identify an evolution from the O8-9 Ia supergiants to the early-B hypergiants. Specifically an emission component appears in the  photospheric profile of He\,{\sc i} 2.059$\mu$m in  many of the O7-9 Ia stars (Figs. 2 \& 3)\footnote{cf. narrow emission in LHO74, -141, -143, and -148 and infilling in LHO 46 and 90.}; likewise the narrow emission features in the red wings of both the He\,{\sc i} 2.059$\mu$m and Br$\gamma$ lines  of the O7-8 Ia stars LHO141 and -148 are replicated and amplified in the BHGs LHO96 and -146 (Fig. 6).

\subsection{Luminous blue variables}

Classification of the three cluster LBVs has been discussed in other works (Figer et al. 
\cite{figer98}, Geballe et al. \cite{geballe00} and Mauerhan et al. \cite{mauerhan10b}) and consequently 
 we simply  highlight the overall  similarity of these stars
to other known LBVs (Fig. 7).
 Spectroscopically  the Pistol star bears direct comparison to AFGL 2298 (Clark et al. \cite{clark09b}),
FMM362 to Wra17-96 (Egan et al. \cite{egan}) and G01.120-0.048 to  the lower luminosity 
star G24.73+0.69\footnote{The previously unpublished spectrum was obtained in 2007 May 9 
utilising  the VLT/ISAAC in the SW MRes mode with a 0.3" slit. The data were reduced following the 
methodology described in Clark et al. (\cite{clark09b}).}. 
 While we defer a detailed analysis  of multi-epoch spectroscopy of the Pistol star, FMM362, and G01.120-0.048 to a future work (Najarro et al. in prep.) we emphasise that the  single-epoch study of Najarro et al. (\cite{paco09}) returned physical properties for the Pistol star and FMM362 that are entirely consistent with quantitative analyses of other LBVs (e.g. Clark et al. \cite{clark09b}, Groh et al. \cite{groh09}), with the {\em explicit} exception of 
$\eta$ Car (Hillier et al. \cite{hillier01}). Specifically $T_{eff}\sim11$kK derived for both stars (Najarro et al. \cite{paco09}) would  imply classification as `cool LBVs' following Groh et al. (\cite{groh14}; henceforth Gr14).

The timescale and magnitude of the near-IR 
continuum variability  exhibited by all three cluster LBVs (Glass et al. \cite{glass99}, Mauerhan et al. 
\cite{mauerhan10b}) is also broadly consistent with that of  field LBVs exhibiting so-called `S Dor variability' (Clark et al. \cite{clark09b}, \cite{clark11}). 
Such a  comparison is critical since it is known that LBVs can experience dramatic eruptions
during which their stellar properties greatly diverge from their quiescent values, with the 19th
century eruption of $\eta$ Carina being a case-in-point. Given the above, we may infer that
none of the  Quintuplet cluster LBVs appears to currently be in such a phase.
Likewise the masses of the nebulae associated with  the Pistol star and 
G01.120-0.048 (9.3$M_{\odot}$ and 6.2$M_{\odot}$ respectively; Lau et al. \cite{lau}) are also comparable
 to those associated with AFGL 2298 and  Wra17-96 (Ueta et al. \cite{ueta}, Egan et al. \cite{egan}).
Thus there is nothing to suggest that the mass-loss history of any of these three stars is atypical for an LBV.

Finally, pre-empting Sect. 3.5, the combination of electron scattering wings associated with the strong 
P Cygni He\,{\sc i} 2.059$\mu$m profile and the presence of 
Br$\gamma$ and Mg\,{\sc ii} emission lines results in  a remarkable similarity between LHO71 and the LBV P Cygni (Fig. 1 and 7; Najarro \cite{paco01}), while LHO67 resembles the known high amplitude photometric variable and consequent LBV candidate
GC IRS34W (Fig. 7; Trippe et al. \cite{trippe}, Martins et al. \cite{martins07}).

\subsection{Wolf-Rayet stars}

A total of six WN stars were reported by Fi99a and Li09; we fail to detect any further examples. We re-classify the WN9h stars LHO67 and -71 as WN10h and WN11h respectively; this is 
 on the basis of (i) the lack of He\,{\sc ii} 2.189$\mu$m in both, (ii) the development of Si\,{\sc iii} and Mg\,{\sc ii} emission in the former, and (iii) the strengthening 
of Mg\,{\sc ii} and the absence of Si\,{\sc iv} emission in the latter (Figs. 1 and 7). Following from the above discussion and adopting the nomenclature of Gr14 we might also consider these as {\em candidate} hot LBVs.

We retain the classification of WN9h for  LH0158, noting that preliminary modelling (Najarro et al. in prep.) suggests that its  strong, broad emission lines  arise in a very cool, dense wind (Fig. 7). Spectra of LHO99 and 
qF274 are presented in Fig. 5. Intriguingly, these appear more similar to the WN8-9h stars found within the Arches (Martins
 et al. \cite{martins08}, Clark et al. \cite{clark18}) than LHO158, presumably as a result of lower density winds with higher terminal velocities than that star (although the reason for such a difference is uncertain; Sect. 5.1).
Finally we maintain the WN6 classification of qF353E, although its location 2.5pc from the cluster core casts some uncertainty as to cluster membership (Steinke et al. \cite{steinke}).

Our combined dataset contains observations of 12 WC stars, with analysis of the thirteenth and final star, qF309, reliant on the spectrum provided in Fi99a\footnote{Mauerhan et al. \cite{mauerhan10a} report on a fourteenth WC star that they label X174617.1 and which they claim is close to the Quintuplet, but a lack of coordinates prevents us  verifying this.}. We retain previously published classifications for nine of these stars, slightly amending the classification of LHO19, qF151, qF309, and WR102ca (Table 1). We are unable to identify any WCE stars within the Quintuplet, a point we return to later. 
Dilution of spectral features and/or near-IR ($m_{\rm F205W}$) photometry suggests eight WC stars are associated with hot dust and a further four possibly so, while a lack of photometry precludes a conclusion for qF76 (Tables 1 and A.1).
Finally our re-reduction of the spectra of the Quintuplet proper members (Fig. 9) clearly reveals the presence of the weak emission lines previously identified by Li09 and Najarro et al. (\cite{paco17}) but no features attributable to the putative massive binary companions.

\subsection{A synopsis}
We find the stellar population of the Quintuplet to be significantly more homogeneous than previous studies had suggested. With the exception of the  single O5 I-III star LHO72, all supergiants span a limited range of spectral types from O7-B0 Ia and  define a smooth morphological  sequence (albeit with the {\em caveat} that four O9-B0 Ia appear anomalously faint).
As such our classifications differ from those of Li09, which accommodate a number  of early-mid O stars. The succession of late-O/early B supergiants naturally extends to a cohort  of B0-3 Ia$^+$ hypergiants and onwards into a regime comprising both hot- and cool-phase (candidate) LBVs/WN10-11h stars. The exceptionally rich early-B hypergiant  and LBV/WN10-11h  populations (seven and five stars respectively) appear to be a defining feature of the Quintuplet; in comparison   
only six field early-B HGs are known (Clark et al. \cite{clark12}), with a 
further seven candidates in young obscured clusters\footnote{Three in Wd1 (Clark et al. \cite{clark05a}), and one each in 1806-20  (Bibby et al. \cite{bibby}),  Danks 1 (Davies et al. \cite{davies}), Mercer 30 (de la  Fuente et al. \cite{dlF}) and Cl1813-178 (Messineo et al. \cite{messineo11}).}, while no other cluster contains a comparable cohort of LBVs. A further WN9h star, LHO158 appears morphologically similar to these objects, albeit supporting a denser wind.

Three O7-8 Ia$^+$ hypergiants were identified. In conjunction with two  WN8-9h stars  they appear distinct from other cluster members, instead being  consistent with the population of the Arches  (2-3Myr; Clark et al. \cite{clark18}).

Only one further  WN star is observed in the vicinity of the Quintuplet, the WN6 star qF353E (WN6), while the WC population consists exclusively  of WC8 and WC9 stars, the majority of which appear to be dust-forming massive binaries. No WNE, WCE or WO stars are present, nor are any cool super-/hypergiants, such as those that characterise Westerlund 1 
(Clark et al. \cite{clark05a}).

\longtab{1}{
\begin{longtable}{lccclll}
\caption{The stellar population of the Quintuplet cluster}\\
\hline
\hline
\multicolumn{2}{c}{ID} & RA & Dec & \multicolumn{2}{c}{Classification} & Notes \\ 
qF \# & LHO \# & (h m s)    &     (d m s) & Current  & Revised & \\
\hline
\endfirsthead
\caption{continued.}\\
\hline
\hline
\multicolumn{2}{c}{ID} & RA & Dec & \multicolumn{2}{c}{Classification} & Notes \\
qF \# & LHO \# & (h m s)    &     (d m s) & Current & Revised & \\
\hline
\endhead
\hline
\endfoot
76     &     & 17 46 17.53 & -28 49 29.0 & WC9        & WC9             & [DWC2011] 66, WR 102h \\
134    &     & 17 46 15.24 & -28 50 03.6 & LBV        & LBV             & Pistol Star, [DWC2011] 68, radio source \\
151    &     & 17 46 14.81 & -28 50 00.6 & WC8        & WC8-9(d? +OB?)     & [DWC2011] 73, WR 102e \\
157    &     & 17 46 13.9  & -28 49 59   & $<$B0 I    & -               &  \\
211    & 019 & 17 46 15.85 & -28 49 45.5 & WC8/9d +OB & WC9d (+OB)      & WR 102ha, QX3, MGM 5-3, $P_{\rm orb}\sim850\pm100$d  \\
231    & 042 & 17 46 14.69 & -28 49 40.7 & WC9d + OB  & WC9d (+OB)             & WR 102dc, QR7, QX5, MGM 5-2  \\
235S   & 034 & 17 46 15.18 & -28 49 41.6 & WC8        & WC8(d? +OB?)             & [DWC2011] 12, WR 102g  \\
235N   & 047 & 17 46 15.16 & -28 49 39.4 & WC8        & WC8(d? +OB?)             & [DWC2011] 13, WR 102f, dusty?   \\
240    & 067 & 17 46 15.94 & -28 49 38.1 & WN9        & WN10h           & [DWC2011] 7, WR 102hb, hot LBV?  \\
241    & 071 & 17 46 15.12 & -28 49 36.9 & WN9        & WN11h           & [DWC2011] 6, WR 102ea, QR5, hot LBV? \\
243    & 075 & 17 46 14.11 & -28 49 36.7 & WC9?d      & WC9d (+OB)            & WR 102da, MGM 5-1   \\
250    & 079 & 17 46 15.38 & -28 49 34.5 & WC9d       & WC9d (+OB)            &     \\
251    & 084 & 17 46 14.77 & -28 49 34.2 & WC9d       & WC9d (+OB)             & WR 102dd, MGM 5-4   \\
256    & 099 & 17 46 16.54 & -28 49 32.0 & WN9        & WN8-9ha          & [DWC2011] 11,  WR102i, Arches-like \\
257    & 096 & 17 46 15.15 & -28 49 32.4 & O6-8 Ife   & B1-2 Ia$^+$     & [DWC2011] 9, QR6, QX2    \\
258    & 102 & 17 46 14.30 & -28 49 31.5 & WC9?d      & WC9d (+OB)            & WR 120db, MGM 5-9 \\
270S   & 110 & 17 46 15.09 & -28 49 29.4 & Of/WN?     & B1-2 Ia$^+$/WNLh & [DWC2011] 8, WR 102df, QR4   \\
274    &     & 17 46 17.53 & -28 49 29.0 & WN9        & WN8-9ha          & [DWC2011] 59, WR 102j, Arches-like  \\
276    &     & 17 46 13.4  & -28 49 29   & B1-3 Ia    & -               &   \\
278    & 077 & 17 46 15.13 & -28 49 34.7 & O6-8 Ifeq  & B0-1 Ia$^+$     & [DWC2011] 10   \\
 301   & 143 & 17 46 16.01 & -28 49 21.8 & O7-B0 I    & O9-B0 Ia        & [DWC2011] 72   \\
 307   &     & 17 46 15.5  & -28 49 20   & B1-3 Ia    & -               & Possible early-B hypergiant?  \\
 307A  & 146 & 17 46 15.48 & -28 49 20.1 & O6-8 I f?  & B1-2 Ia$^+$     & [DWC2011] 69  \\
 309   &     & 17 46 17.51 & -28 49 18.5 & $<$WC8     & WC8-9(d? +OB?)            & [DWC2011] 60, WR 102k    \\
 311   &     & 17 46 13.7  & -28 49 20   & B1-3 Ia    & -               &   \\
 320   & 158 & 17 46 14.05 & -28 49 16.5 & WN9        & WN9h            & [DWC2011] 62, WR 102d, QR8 \\
 344   &     & 17 46 16.7  & -28 49 09   & B1-3 Ia    & O7-8 Ia             & CXOGC J174616.6 -284909  \\
 353E  &     & 17 46 11.14 & -28 49 05.9 & WN6        & WN6             & [DWC2011] 64, WR 102c  \\
 358   &     & 17 46 16.6  & -28 49 05   & B1-3 Ia    & -               &   \\
 362   &     & 17 46 17.98 & -28 49 03.5 & LBV        & LBV             &   \\
 381   &     & 17 46 13.45 & -28 48 59.1 & OB I       & B0-1 Ia$^+$/WNLh & [DWC2011] 4, QR9   \\
 406   &     & 17 46 13.85 & -28 48 50.4 & B1-3 Ia    & O7-8 Ia$^+$     & [DWC2011] 65   \\
       & 001 & 17 46 16.74 & -28 49 51.2 & O3-8 Ife   & O7-8 Ia$^+$/WNLh & [DWC2011] 70    \\
       & 016 & 17 46 15.24 & -28 49 46.5 & O8.5-9.7 Iab? f? & O7-8 Ia   &   \\
       & 026 & 17 46 15.07 & -28 49 44.5 & O5-B0 I f? &  O7-8 Ia    &   \\
       & 028 & 17 46 15.28 & -28 49 43.7 & O7-9 Ie    & O7-8 Ia         &   \\
       & 029 & 17 46 14.71 & -28 49 43.2 & O9-B2 If?e & B1-2 Ia$^+$     &   \\
       & 031 & 17 46 15.16 & -28 49 43.0 & O9-B1 If?  & O7-8 Ia      &   \\
       & 033 & 17 46 16.60 & -28 49 42.7 & O9-B3 I-IIf& O9-B0 Ia     & $m_{\rm F205W}\gtrsim13$, cluster member? \\
       & 039 & 17 46 14.43 & -28 49 40.8 & O7-B1 I f? & O7-8 Ia         &   \\
       & 041 & 17 46 15.83 & -28 49 40.8 & O9-B1 If   & O9-B0 Ia        & $m_{\rm F205W}\gtrsim13$, cluster member?  \\
       & 044 & 17 46 14.94 & -28 49 40.6 & O7-9 If?   & O7-8 Ia         &   \\
       & 046 & 17 46 15.21 & -28 49 40.3 & O7-B1 If?  & O9-B0 Ia        &   \\
       & 048 & 17 46 15.06 & -28 49 39.8 & O7.5-9.5 If?& O7-8 Ia        &   \\
       & 050 & 17 46 16.29 & -28 49 39.4 & O7-B1 If   & O7-8 Ia         &   \\
       & 051 & 17 46 15.78 & -28 49 39.2 & O7-9 If    & O7-8 Ia         & blend?   \\
       & 054 & 17 46 14.46 & -28 49 39.0 & O7-9 I-IIf?& O7-8 Ia$^{(+)}$ &   \\
       & 062 & 17 46 16.58 & -28 49 37.9 & O7-B1 IIIf & OB & $m_{\rm F205W}\gtrsim13$, cluster member? \\ 
       & 064 & 17 46 15.59 & -28 49 37.7 & O7-9 If?   & OB      &  $m_{\rm F205W}\gtrsim13$, cluster member? \\
       & 069 & 17 46 14.29 & -28 49 37.3 & O6-9.7 If? & O7-8 Ia         &   \\
       & 072 & 17 46 15.31 & -28 49 36.9 & O4-6 Ieq?  & O5-6 Ib-III     &   \\
       & 073 & 17 46 14.92 & -28 49 36.7 & O6.5-7 If? & O7-8 Ia        &   \\
       & 074 & 17 46 14.50 & -28 49 36.6 & O9.5-B1 Iab?f& O9-B0 Ia      &   \\
       & 076 & 17 46 14.15 & -28 49 35.2 & WC9d       & WC9d (+OB)           & [DWC2011] 71   \\
       & 088 & 17 46 14.29 & -28 49 33.8 & O3-4 IIIf  & OB & $m_{\rm F205W}\gtrsim13$, cluster member? \\
       & 089 & 17 46 15.00 & -28 49 33.2 & O7.5-8.5 If & O9-B0 Ia       &   \\
       & 090 & 17 46 14.88 & -28 49 33.3 & O7-9.5 Ife? & O9-B0 Ia       &   \\
       & 094 & 17 46 15.73 & -28 49 32.6 & O7-B1 If?   & OB & $m_{\rm F205W}\gtrsim13$, cluster member?\\ 
       & 100 & 17 46 15.15 & -28 49 31.4 & O6-8 Ife    & B2-3 Ia$^+$    &   \\
       & 103 & 17 46 14.16 & -28 49 31.5 & O7-9 If     & OB       &  $m_{\rm F205W}\gtrsim13$, cluster member? \\
       & 105 & 17 46 15.58 & -28 49 31.0 & O7-9 If?    & OB         & $m_{\rm F205W}\gtrsim13$, cluster member?  \\
       & 118 & 17 46 16.05 & -28 49 27.4 & O6-9 If?    & O7-8 Ia     &   \\
       & 122 & 17 46 14.21 & -28 49 26.7 & O7-9.7 Ie   & O7-8 Ia        &   \\
       & 128 & 17 46 14.57 & -28 49 25.9 & O3-5 I      & O9-B0 Ia      & $m_{\rm F205W}\gtrsim13$, cluster member?  \\
       & 132 & 17 46 14.40 & -28 49 24.6 & O7-9 I-IIf? & O9-B0 Ia      & $m_{\rm F205W}\gtrsim13$, cluster member?  \\ 
       & 141 & 17 46 14.99 & -28 49 22.3 & O9.7-B1 I   & O7-8 Ia        &   \\
       & 144 & 17 46 15.34 & -28 49 20.4 & O7-9 I      & O7-8 Ia       &   \\
       & 148 & 17 46 15.05 & -28 49 18.9 & O7-9 I      & O7-8 Ia       &   \\
       & 149 & 17 46 14.00 & -28 49 18.6 & O7-9.7 I    & O7-8 Ia        &   \\
       &     & 17 46 13.05 & -28 49 25.1 & WC8-9       & WC9d (+OB)           &  [DWC2011] 63, WR 102ca \\
       &     & 17 46 05.63 & -28 51 31.9 & LBV         & -              & [DWC2011] 92,  LBV G0.120 -0.048 \\
 \end{longtable}
{ Columns 1 and 2 provide the identifier for the star following the nomenclature of Fi99a and
Li09 respectively. Columns 3 \& 4 provide co-ordinates. Column 5 provides the most 
up-to-date spectral classification available from either Fi99a or Li09, while column 6 provides our revision. Finally 
column 7 provides any additional widely used identifiers (e.g. WR number and  the naming convention of Dong et al. 
\cite{dong11}). qF309 and -344 are re-classified via comparison of the  spectra from Fi99a and Mauerhan et al. 
(\cite{mauerhan10a}) to our new data. The five Quintuplet proper members are those stars with MGM 5-\# identifiers 
(Moneti et al. \cite{moneti92}).
QX\# and QR\# indicate radio and X-ray detections reported in Lang et al. (\cite{lang}) and Wang et al.
(\cite{wang}) respectively. The Pistol star is also reported as a radio source but
no new identifier is assigned for it, while the X-ray designation of qF344 derives from Muno et al. (\cite{muno09}).
 Tuthill et al. (\cite{tuthill}) provide the orbital period for qF211 and suggest periods in the hundreds of days for qF243, -251, and -258 based on the size of their dusty nebulae.}}

\begin{figure*}
\includegraphics[width=14cm,angle=-0]{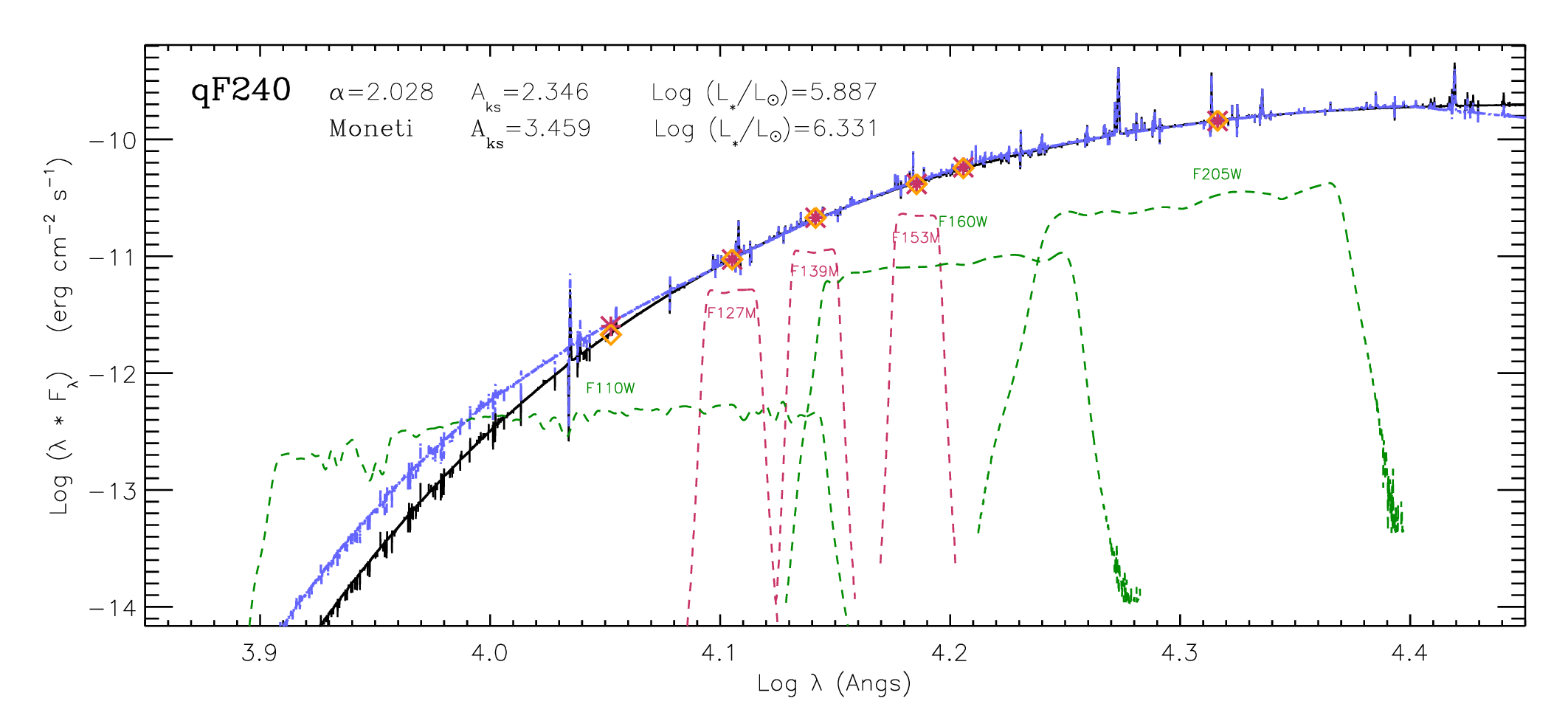}
\includegraphics[width=14cm,angle=-0]{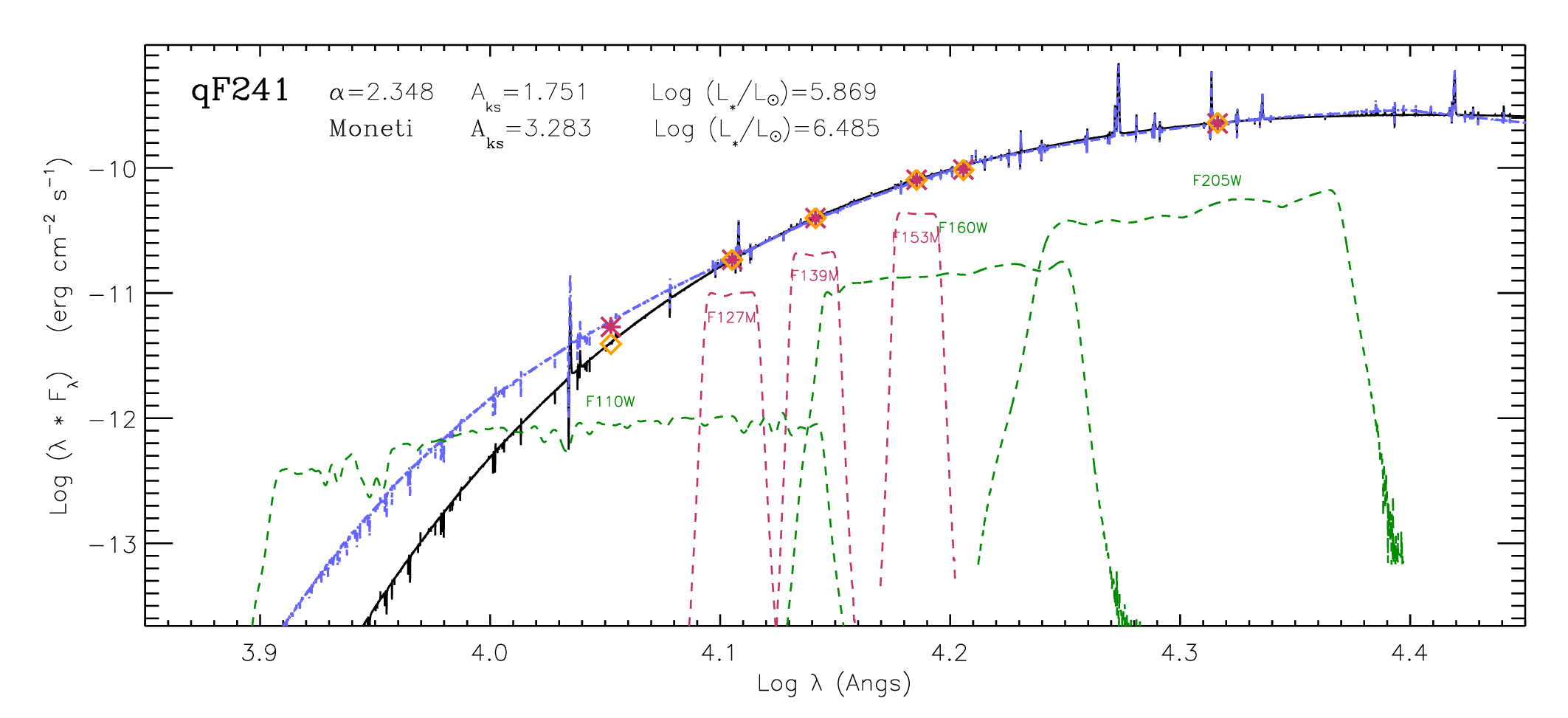}
\caption{Synthetic model-atmosphere spectra for the cluster members qF240 (WN10h) and qF241 (WN11h) computed for two differing assumed interstellar reddening laws, illustrating the dramatic dependence of bolometric luminosity on this choice.
HST photometry employed are from Table A.1. The black lines reflect the model spectra    reddened by $\alpha$=2.028 and 2.348, which results in $A_{ks}$=2.346 and 1.751, respectively. The blue lines follow Moneti's law with
$A_{ks}$ =3.459 and 3.283 respectively. Transmission curves for the filters
 used for the fit are shown in green (broadband) and pink (narrowband), and symbols are plotted for each magnitude
measurement to show the goodness of fit: orange diamonds for the $\alpha$-model and pink stars for the Moneti model.
  The x-axis position of each symbol corresponds to the classical $\lambda_0$ of the filter at which the zero-point
 flux is defined. The y-axis position coincides with its corresponding model curve if the observed magnitude matches
 the magnitude of the reddened model. }
\end{figure*}

\section{Global cluster properties}

As highlighted in Sect. 1, the Quintuplet  and its constituent stars offer the prospect of advancing our knowledge of 
single and binary stellar evolution and the formation of both massive stars and star clusters in extreme environments. In 
order to achieve these goals we must quantify its basic properties, such as age, integrated mass and  
(initial) mass function ((I)MF), which in turn demands an accurate determination of the bolometric luminosities of cluster members
from which (initial) stellar masses and ages may be inferred. However, this requires knowledge of the correct extintion law to 
apply, for which  no current consensus exists,  with Cardelli (\cite{cardelli}), Nishiyama et  al. (\cite{nishiyama}), Moneti et 
al. (\cite{moneti01}), and Hosek et al. (\cite{hosek}) all providing differing prescriptions.

In order to illustrate the difficulties this uncertainty leads to,  
in Fig. 10 we present the results of a preliminary modelling of the spectra and spectral energy distributions  of 
LHO67 (=qF240; WN10h) and LHO71 (=qF241; WN11h) with the model-atmosphere code CMFGEN (Hillier et al. \cite{hillier98}, \cite{hillier99}).
Specifically, following the methodology employed in our analysis of the Arches (Clark et al. \cite{clark18}; 
and for which an identical problem exists)  we employed both a simple power-law model and the   Moneti et al. (\cite{moneti01})
reddening law, which is optimised specifically for the Quintuplet. While both prescriptions provide acceptable fits to the data, 
the resultant bolometric luminosities differed by up to $\sim0.6 (0.45)$ dex for LHO71 (LHO70). As a result of this substantial  
uncertainty - and with the prospect of differential reddening across the cluster field
  - we refrain from the construction of an H-R diagram and consequent comparison to isochrones to determine cluster 
parameters at this time.  We note that this issue afflicts previous analyses (e.g. Liermann et al. \cite{liermann12}) which 
are also prone to additional   uncertainties in both bolometric luminosity (due to a reliance on ground-based photometry) 
and  adopted stellar temperatures (as a result of spectral classification based on data of lower S/N). 

 For the same reasons we also refrain from determining  a cluster luminosity function at this time, from which  
estimates of the (I)MF and integrated  cluster mass could be made. An additional concern is the luminosity/mass function appears 
degenerate at this epoch, with e.g. a number of  the more evolved (and hence presumably initially more massive) WC stars 
being fainter than  the less evolved BHGs, and the WC cohort itself showing $\Delta m_{\rm F205W} \sim 4$mag. One could simply 
choose to adopt a canonical mass function (e.g. Kroupa) but previous studies suggest that the correct form is flatter (Figer 
et al. \cite{figer99b}, Schneider et al. \cite{schneider14}); given these mutiple issue we conclude that any further 
discussion is currently premature.

 Nevertheless, despite the lack of a main-sequence turn-off in our data we may still make provisional determinations of the 
 cluster age and (initial) stellar masses for constituent stars via comparison of the post-MS population to observations of less obscured 
clusters and synthetic stellar spectra derived from theoretical evolutionary calculations.

\subsection{Quintuplet age via comparison to other clusters}

Comparison of the stellar content of the  Quintuplet to other clusters, particularly those for which isochrone fitting
has been possible, provides an empirical age estimate. As shown in Clark et al. (\cite{clark13}) comparatively few clusters 
host WC stars. Of these, while WC8-9 stars are found within Westerlund 1 ($\sim5$Myr; Clark et al. \cite{clark05a}, Crowther 
 et al. \cite{crowther06}) and the Galactic Centre cluster ($\sim6$Myr; Paumard et al. \cite{paumard06}, Martins et al. 
 \cite{martins07} and Feldmeier-Krause et al. \cite{FK}), the presence of cool super-/hypergiants within both marks them as 
 older than the Quintuplet. At the other end of the scale, despite the possible association of a WC7 star, the presence of an O3 
 giant and  supergiant indicates that Pismis 24 is substantially younger than the Quintuplet (1-2Myr; Massey et al. 
 \cite{massey}). At $\sim2-3$Myr  the Arches  cluster provides a more robust lower limit (Clark et al. \cite{clark18}), with the 
 mid-O super/hypergiant population demonstrably earlier than the corresponding late O/early-B stars that dominate the 
Quintuplet (Sect 3), with the notable exception of the small, anomalously early cohort presented in Fig. 5.

 Can we identify any direct comparators? Both the Quartet (4$^{+4}_{-1}$Myr; Messineo et al. \cite{messineo09}) and Danks 2 
 ($3^{+3}_{-1}$Myr; Davies et al. \cite{davies}) have similar stellar contents and ages derived from isochrone fitting which are 
 consistent with the above constraints. However the cluster associated with SGR1806-20 looks to provide the closest counterpart, 
 given it contains an LBV, an early B hypergiant, late O/early B supergiants, two late WN  and two WC9 stars
 (Eikenberry et al. \cite{eikenberry}, Figer et al. \cite{figer05}, Bibby et al. \cite{bibby}). An age of $\sim3-5$Myr, derived 
 from stellar content and isochrone fitting (Bibby et al. \cite{bibby}), is also full compatible with expectations. Moreover its 
 location - only $\sim1.6$kpc from the Galactic centre (Bibby et al. \cite{bibby}) - is of particular interest since it appears 
 to have  formed in a similarly high  metallicity environment to the Quintuplet.

\subsection{Comparison to theoretical predictions}

The output of evolutionary codes  - stellar parameters such as luminosity and temperature - have historically been presented as numerical values tabulated as a function of initial mass and age, from which e.g. isochrones may be constructed (e.g. Ekstr\"{o}m et al. \cite{ekstrom}). Recently Gr14 and Martins \& Palacios (\cite{martins17}; henceforth MaP17) have pioneered a new approach in which these are used as input into the non-LTE model atmosphere code CMFGEN, from which synthetic spectra and absolute magnitudes  are derived in order to permit direct comparison to observables. 

MaP17 present synthetic spectra of non-rotating H-burning stars of solar metallicity for masses between 15-100$M_{\odot}$ which are suitable for comparison to the non-WRs within the Quintuplet. The least evolved object present appears to be the O5 I-III star LHO72. MaP17 find O5 giants only occur in the  $50M_{\odot}$ and $60M_{\odot}$ tracks, with O5 supergiants descending from the most massive ($\geq80M_{\odot}$) progenitors. Moving onto the {\em bona-fide} supergiant cohort and the $50M_{\odot}$ track fails to accommodate the presence of O7-9Ia stars. Conversely, while both the $80M_{\odot}$ and $100M_{\odot}$ tracks predict late O supergiants comparable to those we observe, the range of spectral types extends to much earlier (O3-5) examples than appear to be present at this time. If the Quintuplet is indeed older than the Arches - which does host O4-5 supergiants - such very massive stars would be expected to have already evolved through this 
phase to become WR stars, which were not simulated by MaP17. In contrast the $60M_{\odot}$ track yields a narrower range of spectral types for supergiants (${\sim}$O7.5-9.5 Ia) that is broadly compatible with our data. Finally the simulations suggest that the early B-hypergiants may descend from a wide range of initial masses ($50-100M_{\odot}$), with the widest spread in spectral types ($\sim$B0-0.7 Ia$^+$) found for $60M_{\odot}$ progenitors. These predictions  would be consistent with majority of pre-WR super-/hypergiants within the Quintuplet have evolved from stars of $\sim60M_{\odot}$, with the consequence that the  WR component evolved from more massive progenitors ($\gtrsim80M_{\odot}$).

 In contrast Gr14 present simulations of a single, non-rotating $60M_{\odot}$ star from the zero-age main sequence through to supernova. Fortuitously, such an evolutionary track represents an excellent match for stars within the Quintuplet up until the onset of the H-depleted Wolf-Rayet chapter of their lives. An O7.5 Iafc phase (LHO39?) is first reached  at $\sim3.02$Myr, with the star transitioning from B0.2 Ia to B0.5Ia$^+$ (e.g. LHO143 and LHO77 respectively) after $\sim3.22$Myr. After $\sim0.079$Myr as a BHG there follows an extended period of 0.235Myr during which the star appears as either a hot (e.g. LHO67 and -71) or cold (e.g. the Pistol star and qF362) LBV, before rapidly evolving through a WNLh phase (e.g. LHO58) in only 
5000yr, at which point it is $\sim3.54$Myr old. Subsequently the star presents as WNE, WCE and finally WO before experiencing core-collapse at $\sim4$Myr.

The classifications of at least 38 cluster members are consistent with predictions for this pathway: the single O5-6I-III star, the 18 O7-8 Ia and five bright O9-B0 Ia supergiants (including qF344 in the former grouping), the seven BHGs, the six LBVs and WN9-11h stars and the single  WN6. Moreover it is arguable that the two WN8-9h stars (LHO99 and qF274) and the five stars for which only provisional classifications are available (footnote 9 and Sect. 3.2) could also be accommodated. Another attractive feature of this evolutionary scheme is the comparatively extended period that stars present as either  BHGs or LBVs ($\sim0.314$Myr). As Gr14 notes this is significantly longer than typically assumed for the lifetime of the LBV phase, but helps explain the large number of such stars currently present within the Quintuplet. Similarly a greater duration is predicted for the mid-late O supergiant phase in comparison to the late-O/early B supergiant stage, again consistent with the much larger population of O7-8 Ia stars compared to the bright O9-B0 Ia cohort. 

We emphasise that this evolutionary pathway is not synonymous with an isochrone in the sense that one would not expect all spectral types along it to be present simultaneously at any given time. Nevertheless the brevity of the evolutionary phases proceeding from O7 supergiant to WNLh star - $\sim15$\% of the total lifespan of a $60M_{\odot}$ star - suggests that we might reasonably expect examples of each of these distinct spectral subtype to be present, especially if (i) the cluster comprises stars with a range of rotational velocities and (ii) the evolutionary pathways of stars with masses slightly higher or lower than $60M_{\odot}$ do not deviate substantially from the predictions of Gr14 (see below).

However, discrepancies between the empirical spectral classifications and theoretical predictions are  present. 
The schemes of  Gr14 and MaP17  fail to replicate the three late-O hypergiants (LHO1, -54, and qF406) as well as the second, fainter subset of O9-B0 supergiants observed. Taken at face value the H-free WR population also appears discrepant; while a single WN6 star (qF353E) may be associated with the Quintuplet, the WNE and WCE populations predicted by Gr14 appear absent; conversely Gr14 fail to forecast the rich population of WCL stars within the Quintuplet. 
Being more evolved than the O super-/hypergiants, the WR population must have evolved from stars with masses $>>60M_{\odot}$.
Unpublished simulations indicate that WRs derived from  such  massive stars also yield  WNE and WCE 
spectral subtypes (Jose Groh, 2018, priv. comm.); hence we may not appeal to differing evolutionary pathways at very high masses in order to explain their absence from the Quintuplet.

Nevertheless, despite these tensions we suggest  that the Quintuplet  has an  age of 
$\sim3-3.6$Myr  with stellar masses for the  late-O/early-B super-/hypergiants and LBVs in excess of $50M_{\odot}$ and most likely around $\sim60M_{\odot}$. Unfortunately the difficulty in reconciling the observed WR population to theoretical predictions  prevents us from inferring an age for this cohort, although they must have evolved from very massive stars. Indeed at such an age it is highly likely that some stars have already been lost to SNe, with Groh et al. (\cite{groh13}) suggesting core-collapse for $120M_{\odot}$ ($85M_{\odot}$) stars after 3.0Myr (3.4Myr).

\subsection{Corroboration - luminosities and masses of cluster members}

While the non-rotating $60M_{\odot}$ evolutionary channel explored by Gr14 impressively reproduces
the pre-WR stellar subtypes present within the Quintuplet, it also makes further predictions regarding the masses and luminosities
of such stars. Are these also realised? To date no eclipsing binaries have been identified within the Quintuplet, preventing 
dynamical mass determinations. However in conjunction  with theoretical/spectroscopic modeling we may utilise values derived from known examples external to the Quintuplet  to predict masses for cluster members as a function of spectral type and luminosity class.

Turning first to cluster members and Najarro et al. (\cite{paco09}) present a detailed
quantitative spectroscopic analysis for the Pistol star and FMM362, finding, for an assumed
$A_K\sim3.2$mag,   bolometric luminosities of  $L_{\rm bol}\sim1.6\times10^6L_{\odot}$ and $1.8\times10^6L_{\odot}$, respectively. Our preliminary 
 modelling results for the closely related WN10h and WN11h stars LHO67 and LHO71 
yield luminosities of $L_{\rm bol}\sim(0.8 -3.0)\times10^6L_{\odot}$ for the range
of possible extinctions considered ($A_{\rm Ks}\sim1.8-3.5$; Fig. 10)\footnote{For comparison the independent analysis of these 
stars by Liermann et al. (\cite{liermann10}) adopting the Moneti et al. reddening law placed them at the upper limits of our 
luminosity range and returned moderately higher temperatures.}. These estimates show all four stars to be of comparably high luminosity; by comparison Gr14 predict luminosities of 
$L_{\rm bol}\sim0.8(1.3)\times10^6L_{\odot}$ for hot(cool) LBVs, in excellent agreement with our values. Based on the $V_{\infty}/V_{\rm esc}$ ratio Najarro et al.
(\cite{paco09})  return current masses of $27.5M_{\odot}$ and
$46M_{\odot}$ for the Pistol Star and FMM362 respectively. Despite the inevitable uncertainties in these values as a result of the treatment of reddening, they are fully consistent with the range of masses that the simulations of Gr14 suggest are spanned by LBVs.

The only other star for which a luminosity estimate is available is the WC9d(+OB) star LHO19 (=qF243, MGM 5-3), for which Najarro et al. (\cite{paco17}) derive $L_{\rm bol}\sim1.2(2.8)\times10^5L_{\odot}$ for the WR primary for an assumed $A_K \sim 2.54(3.3)$
(Nishiyama et al. \cite{nishiyama} and Moneti et al. \cite{moneti01} prescriptions). Despite the  lower luminosity derived for LHO19 being  consistent with the brightest
WC9 field stars studied by Sander et al. (\cite{sander}), LHO19 appears less luminous than the WC and WO stars predicted by Gr14 to result from the $60M_{\odot}$ evolutionary pathway (we return to this topic in Sect. 5.2).

One may also ask whether extant dynamical mass estimates for field stars  validate the predictions of Gr14. We are only aware of three direct determinations of O supergiant masses. Of these the primary of V729 Cyg (O7 Ianfp, $\sim31.9\pm3.2M_{\odot}$; Linder et al. \cite{linder}) and the secondary in Wd1-13 (O9.5-B1 Ia, $\sim35.4\pm5M_{\odot}$; Ritchie et al. \cite{ritchie}) are both lower than the $\sim50M_{\odot}$ reported by Gr14 although there is some doubt as to the applicability of both as benchmarking systems\footnote{The lightcurve fitting that constrains the inclination of V729 Cyg yields a distance significantly less than the canonical 1.4kpc assumed for the Cygnus OB2 association, resulting in some uncertainty as to the reliability of the analysis, while the evolution of the secondary in Wd1-13 has clearly been modified by binarity.}.
Mass estimates for both super- and hypergiant components of Cyg OB2 B17 (O7 Iaf$^+$  + O9 Iaf, $60{\pm}5 + 45{\pm}4M_{\odot}$; Stroud et  al. \cite{stroud}) are, however, fully consistent with expectations, although we caution that binary interaction may have already occurred in this system. A
lower limit of $39M_{\odot}$ is inferred for the B1 Ia$^+$ primary BP Cru in the non-eclipsing X-ray binary GX301-2,  with an upper limit of  $\sim53(68)M_{\odot}$ for a neutron star mass of 
$\sim2.5(3.2)M_{\odot}$ (Kaper et al. \cite{kaper}); again, subject to binary interaction, consonant with Gr14.
Finally both stars within  the eclipsing contact binary GC IRS16SW  
($2\times$Ofpe/WNLh; Martins et al. \cite{martins06b}) share similar spectral morphologies to  LHO110.   Peeples et al. (\cite{peeples}) derive dynamical  masses  of $\sim50M_{\odot}$ for both components of GC IRS16SW; compatible with Gr14 if they  are indeed evolving from a B-hypergiant to an LBV phase.

\section{Discussion}

To summarise the above discussion -  excluding the H-free WR phase we find an excellent correspondence between the classification of the majority of massive stars within the Quintuplet and theoretical predictions for non-rotating, single stars of $60M_{\odot}$ and solar metallicity after $\sim3-3.6$Myr of their lives. This is  corroborated by (limited) extant observational data on comparable clusters and `field' stars with similar classifications to those within the cluster. Nevertheless further constraints would be invaluable; particularly quantitative modelling of cluster members subsequent to the determination of the appropriate reddening law and/or the identification of an eclipsing binary to calibrate the cluster mass/luminosity relation.

An important {\em caveat} is that the theoretical models employed  to construct both  evolutionary pathways as a function of stellar mass and cluster isochrones assume single star evolution  - if instead the 
Quintuplet is dominated by a binary channel then results derived from them will be in error.  
Unfortunately no systematic radial velocity survey for binarity has been attempted for the Quintuplet.
Nevertheless, radial velocity surveys of massive stars suggest rather high binary fractions  
(e.g Sana et al. \cite{sana12}, \cite{sana13}) - leading to the expectation of interactions amongst  a sizeable percentage 
that will in turn substantially modify their evolutionary pathways (de Mink et al. \cite{demink13}, \cite{demink}).
Further motivated by the possible non-coevality of the Quintuplet (e.g. Liermann et al. \cite{liermann12}), Schneider et al. (\cite{schneider14}) investigated the impact of binarity on the integrated cluster properties, suggesting that they were consistent with a coeval cluster of  $4.8\pm1.1$ Myr and  a binary fraction of 60\%, where the $8\pm3$ most 
luminous/massive  objects - such as the Pistol star -  are binary products.

However, with the downwards revision of the luminosity of the Pistol star with respect to earlier determinations (Najarro et al \cite{paco09}) and our revised stellar classifications (Sect. 3), it is no longer clear that the Quintuplet shows a significant degree of non-coevality (Sect. 4.2). Moroever under the evolutionary scheme of Schneider et al. (\cite{schneider14})  one might anticipate that a subset of the  $\sim40$\% assumed single stars would have evolved into highly luminous cool super-/hypergiants  if the cluster were as old as  $\sim$4.8Myr, but none are apparent (cf. Sects. 3.1 and 4.1). Finally, the conclusions of Schneider et al. (\cite{schneider14}) rely on fitting the K-band cluster luminosity function, but our findings for both the Arches and Quintuplet clusters suggest that the errors they attribute to this may have been underestimated given both the uncertainty in the correct extinction law to apply and the possibility of significant differential reddening.

Notwithstanding these findings, even if the bulk of the pre-WR population of the Quintuplet can adequately be explained via a combination of a single star evolutionary channel and the younger cluster age proposed in Sect. 4, there exist three sub-populations within the Quintuplet that are difficult to accommodate in such a scenario.

\subsection{Late-O hypergiants and WN8-9h stars}

The first subset of objects that diverge from the predictions of Gr14 are the O hypergiants LHO1, -54, and qF406 and the  WN8-9h stars LHO99 and qF274 (Fig. 5). Gr14 predict a short post-LBV WNL(h) phase which is {\em potentially} consistent with the presence of LHO99 and qF274. However their spectral morphologies differ from other  WN9-11h cluster members (compare to LHO67, -71, and -158 in Fig. 7) which seem to support much slower, denser  winds - one might ask why this is the case. Critically however, an O hypergiant phase is not predicted by Gr14 for $60M_{\odot}$ stars. Likewise the fact that LHO1 is significantly brighter than LHO54 and the remaining O7-8 supergiants means it is difficult to attribute its differing appearance solely to an anomalously dense wind.

Intriguingly all five stars appear strikingly similar to the more massive members of the younger ($\sim2-3$Myr) Arches cluster. Indeed, a close evolutionary relationship between mid-late O hypergiants and WN8-9h stars is suggested by  the stellar population of this aggregate (Clark et al. \cite{clark18}), implying that LHO1, -54, and -99 and qF274 and -406 could represent a physically coherent sub-population. 
 Does this similarity mean that these stars are younger and more massive than the remaining members of the  Quintuplet, which would therefore not be co-eval, or that we are looking at two distinct stellar populations along the same line of sight?  While proper motion measurements would help distinguish between these hypotheses, we can also anticipate two further possibilities. 

Firstly, evolutionary codes (e.g. Ekstr\"{o}m et al. \cite{ekstrom}, de Mink et al. \cite{demink09}) show that very massive and rapidly rotating stars do not evolve redwards across the HR diagram, instead remaining at high temperatures until core-collapse. Recent studies of the distribution of the rotational velocities of massive stars (Ram\'{i}rez-Agudelo et al. \cite{RA}, Dufton et al. \cite{dufton}) suggest distinct slowly- and rapidly-rotating populations. One might therefore suppose that LHO1, -54,  -99 and  qF274 and -406 are rapid-rotators and, as a consequence, are evolving along an alternative pathway to the remaining cluster population (which we would infer to be slow rotators). 
While we see no obvious evidence for rapid-rotation in the current data we may not yet fully exclude this possibility since we might be viewing the stars under an unfavourable inclination and/or spin-down may have already occurred.
 The origin of such a {\em putative} cohort of rapid-rotators is uncertain, but it has been suggested that such stars may represent the  secondaries in binary systems that have been spun-up via mass-transfer (de Mink et al. \cite{demink13}). 

Alternatively, we highlight that Westerlund 1 hosts similarly distinct populations of both early (B0-1; Wd1-5, -13, and -44)
and late (B5-9; Wd1-7, -33, and -42) hypergiants (Clark et al. \cite{clark05a}, Negueruela et al. \cite{negueruela}). Observationally, the former either currently  reside in, or are inferred to have belonged to, massive binaries (Ritchie et al. \cite{ritchie}, Clark et al. \cite{clark14b}, in prep.), while no evidence for binarity is found for the latter stars. In contrast to the previous scenario,  Wd1-5, -13, and -44 are thought to be the stripped primaries resulting from binary-driven mass-loss, while Wd1-7, -33, and -42 have evolved via a single star channel. We might therefore naturally assume that LHO1, -54, and qF406 also result from binary-driven mass stripping. 

Since both scenarios implicate binarity as a physical driver, a multi-epoch radial-velocity survey and detailed quantitative analysis of these  stars in order to identify signatures of binary-modified evolution (e.g. over-luminosity and/or anomalous chemical abundances; cf. Wd1-5, Clark et al. \cite{clark14b}) would be of considerable interest.

\subsection{The hydrogen-free Wolf Rayet cohort}

The main point of divergence between our observations and theoretical predictions for very massive stars  is the nature of the H-free Wolf Rayet population. Specifically, Gr14 predicts WN2-5 and WC4 phases for  $60M_{\odot}$ stars, with unpublished simulations predicting identical WNE and WCE phases for  both $85M_{\odot}$ and $120M_{\odot}$ stars (Jose Groh, 2018, priv comm.); all three pathways also predict a final WO phase immediately before core-collapse (Groh et al. \cite{groh13}). 
Instead we find a lone, apparently H-free, WN6 star in close proximity to the Quintuplet (qF53E; Steinke et al. \cite{steinke}) and  no WCE or WO stars, although a rich population of WCL stars is present (Table 1).

Given the apparent age of the Quintuplet,  WO stars originating  from the most massive progenitors ($\geq85M_{\odot}$; Groh et al. \cite{groh13}) might be expected to be present. The apparent lack of such stars is most easily explained via the brevity of this phase and predictions for absolute magnitudes some $\sim2-2.5$mag fainter than the O7-8 Ia cohort (Groh et al. \cite{groh13}, Gr14); they are likely too faint to be detected. However neither factor can explain the lack of WNE stars, which the $60M_{\odot}$ evolutionary track of Gr14 suggests should be of comparable  magnitude to the O7-8 supergiants. WCE stars evolving from $60M_{\odot}$ stars  are predicted to be a magnitude fainter than this; nevertheless our observations should also reach 
them (cf. Fig. A.2). Moreover, if present, they  should be easily identifiable via their strong emission lines. As a consequence it appears difficult to understand why we do not see such stars if, as predicted,  $\geq60M_{\odot}$ stars all evolve through WNE and WCE phases. 

One possible hypothesis for the prevalence of WNL and WCL  subtypes over WNE and WCE  is that it is the result of the metallicity of the environment in which they formed - as
suggested by the observational finding that late sub-types  are preferentially observed in the inner, high metallicity, regions of galaxies (e.g. Conti \& Vacca \cite{conti} and Hadfield \& Crowther \cite{hadfield}). For WN stars a higher intrinsic N-fraction naturally biases stars to later spectral subtypes, as does a metallicity-dependent mass-loss rate leading to  stronger, higher density winds for both WN and WC stars (Crowther \cite{crowther07}). Gr14 utilise calculations for solar-metallicities - if instead the CMZ is super-solar then this could contribute to the lack of WNE and WCE stars within the Quintuplet. The population of massive field stars in the CMZ (Mauerhan et al. \cite{mauerhan10a}, \cite{mauerhan10c}, \cite{mauerhan15}) and the older Galactic Centre cluster (Paumard et al. \cite{paumard06}, Martins et al. \cite{martins07}) is broadly consistent with this finding and assertion, with two H-free WN5b being the earliest detected to date and all WC stars being of sub-type WC8-9 with only one exception (the WC5-6 star IRS 3E).

An alternative/complementary explanation is that the single star evolutionary models are deficient in some manner. One might envisage 
several possibilities. Current simulations indicate the temperature at the base of the wind (e.g. the hydrostatic layer) is sufficiently high that 
WCEs are favoured over WCLs; inflation of the outer envelope would help counteract this. Alternatively, as mentioned above, increasing the mass-loss rate would result in a moderate shift to later spectral types, although it is not clear if this would be sufficient to replicate observations (Jose Groh, 2018, priv. comm.). One could also wonder whether the mass-loss prescription for the LBV phase - after which point model and observations appear to diverge - might be incorrect due to the occurence of extensive impulsive  events (as suggested by the nebulae associated with the Pistol star and G01.120-0.048; Sect. 3.4).

 Finally  the WC8-9 cohort may form via a binary channel; the presence of hot dust associated with at least eight stars strongly implies a high binary fraction for the WC population of the Quintuplet (a finding also replicated by the WC8-9 population of Westerlund 1; Clark et al. \cite{clark08}). In this context, a systematic RV spectroscopic survey to determine the physical properties of the  binary population in order to ascertain whether they have undergone  prior interaction would be of considerable interest. However such an hypothesis would not explain the lack of WNE and WCE stars, unless the single-star evolutionary channel was not populated due to an extreme binary fraction.

The issue of binarity  is also directly relevant to a determination of the progenitor masses of WC stars. Interpreting the physical properties of field WC9 stars under a  single-star evolutionary scheme,  Sander et al. (\cite{sander}) suggest that WC stars evolve from stars in the $\sim20-45M_{\odot}$ range and are post-RSG objects. This conclusion seems to be strongly in tension with the findings here, where the WC9 cohort must have evolved from initially very massive stars - certainly $>60M_{\odot}$ and most likely $\gtrsim85M_{\odot}$ - but might be reconcilable if a binary formation channel was/is dominant in the Quintuplet and leads to enhanced mass-loss and hence lower mass and luminosity WC stars.

Intriguingly, the sole WC9 cluster member for which parameters have been determined - LHO19 (=MGM 5-3; Najarro et al. \cite{paco17}) - 
appears to be sub-luminous in comparison to theoretical predictions (Sect. 4.3). Based on the calibration of Langer
 (\cite{langer89}) the range of  luminosities  allowed for the WC component of LHO19 corresponds to current masses of 
$\sim10-12M_{\odot}$. This is already significantly lower than the masses predicted for the pre-core collapse endpoints of non-rotating (rotating) stars of initial masses $120M_{\odot}$, $85M_{\odot}$, and $60M_{\odot}$ - $30.7M_{\odot} (18.9M_{\odot})$, $18.5M_{\odot} (26.2M_{\odot})$, and $12.4M_{\odot} (18.9M_{\odot})$ respectively (Groh et al. 
\cite{groh13}) - suggesting an additional source of mass-loss for this system. Although challenging, quantitative analysis of the remaining WC stars (via assessment of the relative  contributions from primary, secondary and circumstellar dust) appears essential for both internal differentiation - i.e. is LHO19 anomalously faint? - and subsequent comparison to theoretical predictions in order to ascertain their formation 
channel(s).

\subsection{The faint O9-B0 Ia stars}

Finally we turn to the four anomalously faint O9-B0 stars - LHO33, -41, -128, and -132 (Sect. 3.2) - and, by extension, likely a number of stars currently assigned a generic OB classification. Comparison to both cluster- and template-spectra suggest these are supergiants; certainly their relative brightness means they must be evolved post-MS stars (cf. the absolute magnitudes of OB main sequence stars presented in Martins \& Plez \cite{martins06a} and Pecaut  \& Mamajek \cite{pecaut}).
MaP17 suggest that late-O/early B supergiants may evolve from stars of initial mass as low as $\sim40M_{\odot}$ (see also Weidner \& Vink \cite{weidner}), which would be expected to be less luminous than the majority population of the Quintuplet. This  prediction is borne out by observations of Westerlund 1, where the eclipsing binary Wd1-13 suggests initial masses of $35-40M_{\odot}$ (Ritchie et al. \cite{ritchie}) for the rich O9-B3 Ia population (Negueruela et al. \cite{negueruela}, Clark et al. in prep.). However this would imply a significantly greater age for this cohort  in comparison to the remaining population of the Quintuplet, with Westerlund 1 thought to be       $\sim5$Myr old (Negueruela et al. \cite{negueruela}). 

Rapid rotation and/or binary interaction cannot explain this discrepancy, since it would produce a cohort of stars that appear younger, hotter, and potentially more luminous than the dominant population of the Quintuplet,
in direct contrast to observations. As a consequence does this subset of stars indicate that the cluster is indeed non-coeval? While such a conclusion cannot be ruled out - arising either from the merger of sub-clusters or genuine multi-generational star formation in a single stellar aggregate - other options suggest themselves. Embedded within the G305 star-forming complex (Clark \& Porter \cite{clark04}) and separated by a projected distance of $\sim4$pc, the young massive clusters  Danks 1 and 2 show a clear difference in ages ($1.5^{+1.5}_{-0.5}$Myr and $3^{+3}_{-1}$Myr,  respectively; Davies et al. \cite{davies}); one could easily imagine a situation in which similarly  physically distinct clusters  were projected against each other along the same line of sight to the Quintuplet. Intriguingly, Steinke et al. (\cite{steinke}) report the discovery of a handful of OB stars in proximity to the WN6 star qF353E, which they suggest may represent a distinct stellar cluster, {\em  potentially} supportive of the idea of a spatially and  temporally extended episode of star formation that could yield the requisite spread in stellar ages along the line of sight to the Quintuplet.

Alternatively, the CMZ hosts an additional population of apparently isolated massive stars of uncertain origin, which includes late-O/early-B supergiants comparable to these stars  (Mauerhan et al. \cite{mauerhan10a}, \cite{mauerhan10c}). In principle, one could envisage the chance superposition of a number of such `field' stars onto the Quintuplet, although the compactness of the cluster and the homogeneity of the putative population of interlopers  would mitigate against this possibility. In any event determining the origin of this stellar population would seem key to determining the co-evality (or otherwise) of the Quintuplet and the star formation   history in its vicinity.

\section{Conclusions and future perspectives}

 This paper reports the results of a combined photometric (HST/NICMOS+WFC3) and spectroscopic study of the Quintuplet 
 cluster, the latter comprising both archival (VLT/SINFONI) and new (VLT/KMOS) data. Photometry is supplied for   108 objects 
 derived from the combined target lists of Fi99a and Li09a and from which stars of late spectral type have been excluded. We 
 present new spectra of 63 unique objects for which our re-reduction of the VLT/SINFONI data has resulted in a much improved S/N 
 ratio, permitting reliable assessment of faint classification diagnostics such as He\,{\sc ii} 2.189$\mu$m and C\,{\sc iv} 
 2.079$\mu$m. These are supplemented by lower S/N and resolution  published data for a further eight stars, yielding  a 
 final spectroscopic dataset of 71 stars. Analysis of these data results in the re-classification of $\sim70$\% of the  cluster 
 members.

 The major finding of this analysis is that the Quintuplet appears to be far more homogeneous than previous studies have 
 suggested. All supergiants classified were of  spectral types O7-8 and O9-B0 with the exception of a sole O5 I-III star. 
 Despite the suggestion that stars with spectral type as early as O3 might be present (Li09), this is currently the earliest 
 spectral type identified. In terms of  spectral morphology the supergiant cohort smoothly extends through  populations of 
 early-B hypergiants and LBV/WN9-11h stars. However a `parallel channel' comprising mid-late O hypergiants and WN8-9ha stars 
 is also present. No further examples of H-free WRs were detected; these comprise a single WN6 star and a rich population 
 of WC8-9 stars, of which a majority appear to be binaries. No main sequence objects were identified, presumably due to the 
 limited depth of the current observations.

 Due to uncertainty regarding the correct extinction law to employ and the probability of differential reddening, it was not 
possible to quantitatively determine a  cluster age via isochrone fitting.  We likewise refrain from determining a 
cluster mass function and hence estimating an integrated cluster mass at this time, noting simply that the large 
number of hitherto rare spectral sub-types (e.g. BHGs and LBVs)  attests to the presumably extreme mass of the Quintuplet.
Nevertheless, comparison to other young massive clusters and the output of a combination of 
{\em single-star} evolutionary and spectral synthesis codes suggests a cluster  age of $\sim3.0-3.6$Myr. This implies progenitor masses of 
$\sim60M_{\odot}$ for the majority of cluster members that have yet to reach the H-free WR phase ($\geq38$ stars; Sect. 4.2), with that cohort presumably evolving from more massive stars still.

While the vast majority of cluster members  appear entirely co-eval, a handful of objects  potentially challenge this conclusion for the Quintuplet as a whole. The five mid-late O hypergiants and WN8-9ha stars look somewhat younger, their appearance being  consistent with membership of the 2-3Myr old Arches cluster. Potential explanations for this discrepancy include the effect of rapid rotation, possibly
induced by mass-accretion in an interacting binary, or binary-driven mass-stripping  (cf. Wd1-5; Clark et al. \cite{clark14a}). This would be consistent with the predictions of e.g. van Bever \& Vanbeveren (\cite{vbvb}) and Schneider et al. (\cite{schneider14}) that binary products should be present within the Quintuplet.
We also identify a small population of underluminous O9-B0 Ia stars which appear to represent an older population of lower initial masses (potentially also revealed by the RSG  LHO7). Given the dense stellar environment of the CMZ, we cannot at present distinguish between the Quintuplet being non-coeval or 
the chance superposition of interlopers - whether a coherent physical aggregate or isolated field stars  - along the cluster sight line. A combination of proper-motion and radial-velocity data would help discriminate between these possibilities.

Nevertheless, the combination of the age and mass of the Quintuplet presents unique opportunities. No other cluster is so richly populated by the closely related BHGs, LBVs, and WN9-11h stars which, as a consequence of their quiescent and impulsive mass loss rates, appear pivotal to massive stellar evolution. Recently, it has been suggested that such stars form via binary interaction and as a result of either dynamical ejection or SNe kicks are preferentially found in isolation rather than in clusters (Smith \& Tombleson \cite{smith}). Rosslowe \& Crowther
(\cite{rosslowe}) have already challenged the ideas that LBVs are more isolated than other massive stars such as WRs and are not located in clusters. The results presented here are further in tension with Smith \& Tombleson (\cite{smith}); in the Quintuplet the LBVs and WN9-11h stars form a smooth evolutionary sequence with the less evolved OB supergiants and B hypergiants (cf. Fig. 1 and Sect. 4.2), with no suggestion that binarity mediates this procession. If present, mass-gaining binary products appear more likely to be found amongst the handful of O hypergiants (and their close spectral relatives the WN8-9ha stars), for which progenitor stars have not been identified. And while the dusty WC9 stars could represent  stripped primaries in interacting binaries, none show evidence of an LBV secondary (nor the LBVs/WN9-11h stars a WR companion).

The H-free WR population of the Quintuplet is also of particular interest. Prior to this phase the single-star $60M_{\odot}$  evolutionary channel (Ekstr\"{o}m et al. \cite{ekstrom}, Gr14) provides an excellent fit to observations:\newline

 O7-O8 Ia $\rightarrow$ O9-B0 Ia $\rightarrow$ B0-3 Ia$^+$ $\rightarrow$ LBV $\leftrightarrow$ WN9h-11h \newline

However as detailed in Sect. 5.2 it appears likely that observations diverge from theory after this point in three distinct ways: the presence of uniformly late WR spectral sub-types; the extreme ratio (13:1) of H-free WC stars relative to WN (discussed by e.g. Li09, Liermann et al. \cite{liermann12}); the fact that the current mass inferred for the WC9 primary of LHO19 appears unexpectedly low
(cf. Groh et al. \cite{groh13}). A number of non-exclusive solutions suggest themselves, including observational limitations/bias, the effect of a possible non-solar metallicity in the CMZ and  potential deficiencies in the single-star evolutionary physics adopted, such as the treatment of mass-loss and its prevalence in comparison to a binary channel. In relation to the final suggestion, the presence of hot dust associated with  a minimum of eight WC stars is indicative of a very high binary fraction amongst this cohort, emphasising the 
potential role of binary interaction in the production of the WCL stars (cf. Schneider et al. \cite{schneider14}), although in isolation it would appear unable to explain the absence of the WCE stars predicted by the  single star evolutionary channel.

What are the future prospects for exploiting the potential of the Quintuplet? In observational terms a multi-epoch RV spectroscopic survey of cluster members to determine the binary fraction of the cluster - and hence the applicability of single-star evolutionary predictions - appears essential. Summation of multiple spectra (cf. Clark et al. \cite{clark18}) would also allow classification of fainter cluster members such as main sequence objects and, if present, intrinsically fainter WR stars such as the WO subtype predicted to immediately precede core-collapse.

In conjunction with this, individually tailored quantitative non-LTE model-atmosphere analysis of cluster members would allow the construction of an HR diagram, from which a cluster age could be inferred. This would also serve to calibrate the cluster (I)MF, from which a cluster mass and  density could be determined, as well as the integrated radiative
and mechanical feedback into the CMZ from the Quintuplet. Most importantly, such efforts would help refine the input physics of stellar evolution codes (the results of which are utilised in cluster spectral-synthesis codes such as Starburst99) and constrain the influence of binarity and the properties of the final pre-SN stellar endpoint.

In particular we highlight the importance of these goals in understanding the formation of the coalescing high mass black hole binaries identified via gravitational wave observations. With an age of 3-3.6Myr we would expect the most massive  stars ($\geq85M_{\odot}$)  within the Quintuplet to be undergoing SNe at this time, which implies that their immediate progenitors should also be present. Groh et al. (\cite{groh13}) suggest these progenitors should have core-masses of $18.5-30.7M_{\odot}$ and yet the mass of the WC9 primary of LHO19 is already substantially lower at $\sim10-12M_{\odot}$. If the remaining population of WCL stars within the Quintuplet are also of comparably 
low mass, it would raise important questions. If stars of $\geq85M_{\odot}$ are unable to form black holes with  masses substantially in excess of $10M_{\odot}$, how massive must their progenitors be? Do they need to form via a different mechanism (e.g. chemically homogeneous evolution; de Mink et al.  
\cite{demink09})? Is their formation even possible  in the local, high metallicity Universe?

\begin{acknowledgements}
Based on observations collected at the European Organisation for Astronomical Research in the 
Southern Hemisphere under ESO programmes  077.D-0281 and   093.D-0306. This research was supported by the Science and
Technology Facilities Council. FN acknowledges financial support through Spanish grants 
ESP2015-65597-C4-1-R and ESP2017-86582-C4-1-R (MINECO/FEDER).
We thank Chris Evans for his invaluable support in preparing for the KMOS observations, and Jose Groh for 
informative discussions.

\end{acknowledgements}

{}

\appendix

\section{Additional spectra and HST photometry}

\longtab{1}{
\begin{longtable}{lccccccccc}
\caption{The stellar population of the Quintuplet cluster}\\
\hline
\hline
\multicolumn{2}{c}{ID} & RA & Dec & $m_{\rm F110W}$ &$m_{\rm F160W}$ & $m_{\rm F205W}$& $m_{\rm F127M}$&$m_{\rm F139M}$ &$m_{\rm F153}$\\  
qF \# & LHO \#         &    &     & (mag) &(mag) & (mag)&(mag) &(mag) &(mag) \\
\hline
\endfirsthead
\caption{continued.}\\
\hline
\hline
\multicolumn{2}{c}{ID} & RA & Dec & $m_{\rm F110W}$ &$m_{\rm F160W}$ & $m_{\rm F205W}$& $m_{\rm F127M}$&$m_{\rm F139M}$ &$m_{\rm F153}$\\
qF \# & LHO \#         &    &     & (mag) &(mag) & (mag)&(mag) &(mag) &(mag) \\
\hline
\endhead
\hline
\endfoot
  134  &   -   & 266.56352 & -28.83425 & 12.69$\pm$0.06 &  9.34$\pm$0.04 &  7.74$\pm$0.02 & 11.79$\pm$0.01 & 10.84$\pm$0.01      &  -   
\\   
  151  &   -   & 266.55717 & -28.82204 & 16.67$\pm$0.08 & 12.97$\pm$0.05 & 11.27$\pm$0.02 & 15.55$\pm$0.01 & 14.40$\pm$0.01     & 
13.38$\pm$0.01  \\ 
  157  &   -   & 266.55780 & -28.83302 &    -     & 13.58$\pm$0.05 & 11.85$\pm$0.02 & 16.62$\pm$0.03 & 15.35$\pm$0.02     & 
14.06$\pm$0.02  \\ 
  211  &  019  & 266.56612 & -28.82927 & 15.36$\pm$0.04 & 10.52$\pm$0.04 &  7.57$\pm$0.05 & 14.53$\pm$0.01 & 13.03$\pm$0.01     & 
11.46$\pm$0.01  \\  
  231  &  042  & 266.56129 & -28.82794 & 14.63$\pm$0.06 &  9.67$\pm$0.03 &  6.80$\pm$0.04 & 14.33$\pm$0.01 & 12.82$\pm$0.01 & 
11.20$\pm$0.01  \\ 
  235S &  034  & 266.56321 & -28.82820 & 17.00$\pm$0.06 & 13.50$\pm$0.05 & 11.73$\pm$0.02 & 15.92$\pm$0.01 & 14.99$\pm$0.02     & 
13.95$\pm$0.01  \\
  235N &  047  & 266.56315 & -28.82760 & 17.07$\pm$0.06 & 13.54$\pm$0.04 & 11.74$\pm$0.02 & 16.06$\pm$0.01 & 14.85$\pm$0.02     & 
14.13$\pm$0.02  \\ 
  240  &  067  & 266.56642 & -28.82713 & 15.05$\pm$0.10 & 11.39$\pm$0.07 &  9.70$\pm$0.02 & 13.90$\pm$0.01 & 12.84$\pm$0.01     & 
11.82$\pm$0.01  \\ 
  241  &  071  & 266.56299 & -28.82693 & 14.33$\pm$0.08 & 10.84$\pm$0.05 &  9.21$\pm$0.03 & 13.18$\pm$0.01 & 12.16$\pm$0.01     & 
11.11$\pm$0.01  \\ 
  243  &  075  & 266.55890 & -28.82682 & 16.80$\pm$0.05 & 11.48$\pm$0.05 &  8.33$\pm$0.02 & 15.92$\pm$0.01 & 14.29$\pm$0.01     & 
12.57$\pm$0.01  \\ 
  250  &  079  & 266.56417 & -28.82623 & 16.15$\pm$0.14 & 12.13$\pm$0.09 &  9.95$\pm$0.06 & 15.34$\pm$0.01 & 14.07$\pm$0.01     & 
12.78$\pm$0.01  \\ 
  251  &  084  & 266.56165 & -28.82616 & 15.94$\pm$0.05 & 11.13$\pm$0.05 &  8.27$\pm$0.02 & 15.12$\pm$0.01 & 13.57$\pm$0.01     & 
12.04$\pm$0.01  \\
  256  &  099  & 266.56895 & -28.82546 & 15.43$\pm$0.05 & 12.13$\pm$0.05 & 10.57$\pm$0.01 & 14.31$\pm$0.01 & 13.37$\pm$0.01     & 
12.45$\pm$0.01  \\
  257  &  096  & 266.56310 & -28.82567 & 15.02$\pm$0.07 & 11.56$\pm$0.05 &  9.90$\pm$0.02 & 13.99$\pm$0.01 & 12.93$\pm$0.01     & 
11.91$\pm$0.01  \\ 
  258  &  102  & 266.55967 & -28.82537 & 18.96$\pm$0.08 & 13.55$\pm$0.04 &  9.78$\pm$0.02 & 17.90$\pm$0.02 & 16.17$\pm$0.02     & 
14.25$\pm$0.01  \\ 
  270S &  110  & 266.56291 & -28.82480 & 14.90$\pm$0.05 & 11.44$\pm$0.06 &  9.86$\pm$0.05 & 13.87$\pm$0.01 & 12.84$\pm$0.01     & 
11.88$\pm$0.01  \\  
  274  &   -   & 266.57305 & -28.82467 & 15.88$\pm$0.07 & 12.43$\pm$0.05 & 10.90$\pm$0.02 & 14.83$\pm$0.01 & 13.85$\pm$0.01     & 
12.86$\pm$0.01  \\ 
  276  &   -   & 266.55593 & -28.82473 & 16.07$\pm$0.09 & 12.56$\pm$0.07 & 10.97$\pm$0.02 & 15.02$\pm$0.01 & 13.97$\pm$0.01     & 
12.97$\pm$0.01  \\ 
  278  &  077  & 266.56301 & -28.82630 & 15.31$\pm$0.09 & 11.85$\pm$0.05 & 10.21$\pm$0.01 & 14.26$\pm$0.01 & 13.18$\pm$0.01     & 
12.21$\pm$0.01  \\ 
  307A &  146  & 266.56450 & -28.82226 & 14.47$\pm$0.05 & 11.05$\pm$0.05 &  9.46$\pm$0.02 & 13.43$\pm$0.01 & 12.40$\pm$0.01     & 
11.42$\pm$0.01  \\ 
  309  &   -   & 266.57294 & -28.82181 & 17.24$\pm$0.07 & 13.60$\pm$0.05 & 11.74$\pm$0.02 & 16.17$\pm$0.01 & 14.97$\pm$0.02     & 
14.24$\pm$0.01  \\
  320  &  158  & 266.55855 & -28.82123 & 16.35$\pm$0.11 & 12.66$\pm$0.04 & 10.82$\pm$0.03 & 15.18$\pm$0.01 & 14.16$\pm$0.01     & 
13.06$\pm$0.01  \\ 
  344  &    -  & 266.56948 & -28.81921 & 17.31$\pm$0.07 & 13.26$\pm$0.04 & 11.47$\pm$0.02 & 16.29$\pm$0.01 & 15.03$\pm$0.02     & 
13.80$\pm$0.01  \\ 
  353E &    -  & 266.54639 & -28.81829 &  -       &         -  &         -  & 16.03$\pm$0.01 & 15.11$\pm$0.02     & 14.12$\pm$0.01  \\ 
  358  &    -  & 266.56899 & -28.81799 & 16.48$\pm$0.08 & 12.45$\pm$0.05 & 10.53$\pm$0.01 & 15.43$\pm$0.01 & 14.16$\pm$0.01     & 
12.93$\pm$0.01  \\ 
  381  &    -  & 266.55605 & -28.81641 &  -         &   -        &         -  & 14.41$\pm$0.01 & 13.30$\pm$0.01     & 12.26$\pm$0.01  
\\
  406  &    -  & 266.55772 & -28.81403 &  -         &   -        &   -        & 15.56$\pm$0.01 & 14.45$\pm$0.01     & 13.40$\pm$0.01  
\\ 
   -   &  001  & 266.56972 & -28.83089 & 15.53$\pm$0.08 & 12.13$\pm$0.07 & 10.61$\pm$0.03 & 14.51$\pm$0.01 & 13.52$\pm$0.01     & 
12.51$\pm$0.01  \\  
   -   &  002  & 266.56656 & -28.83082 & 18.94$\pm$0.08 & 15.70$\pm$0.04 & 14.27$\pm$0.02 & 17.85$\pm$0.02 & 16.90$\pm$0.02     & 
15.97$\pm$0.02  \\  
   -   &  003  & 266.56560 & -28.83082 & 18.27$\pm$0.07 & 14.79$\pm$0.06 & 13.44$\pm$0.03 & 17.17$\pm$0.02 & 16.17$\pm$0.02     & 
15.22$\pm$0.02  \\
   -   &  005  & 266.56307 & -28.83075 & 17.92$\pm$0.08 & 14.36$\pm$0.05 & 12.87$\pm$0.02 & 16.71$\pm$0.01 & 15.69$\pm$0.02     & 
14.72$\pm$0.01   \\
   -   &  009  & 266.56380 & -28.83027 & 14.94$\pm$0.07 & 14.20$\pm$0.05 & 13.95$\pm$0.03 & 14.56$\pm$0.01 & 14.40$\pm$0.01     & 
14.21$\pm$0.01  \\
   -   &  010  & 266.56320 & -28.82992 & 20.10$\pm$0.06 & 16.29$\pm$0.04 & 14.67$\pm$0.01 & 18.82$\pm$0.02 & 17.70$\pm$0.02     & 
16.59$\pm$0.02  \\
   -   &  011  & 266.56845 & -28.82989 & 18.69$\pm$0.05 & 15.41$\pm$0.03 & 14.14$\pm$0.04 & 17.76$\pm$0.02 & 16.81$\pm$0.02     & 
15.93$\pm$0.02  \\ 
   -   &  013  & 266.56933 & -28.82966 & 19.01$\pm$0.07 & 15.64$\pm$0.05 & 14.34$\pm$0.03 & 18.05$\pm$0.02 & 17.05$\pm$0.02     & 
16.08$\pm$0.02  \\ 
   -   &  015  & 266.56406 & -28.82963 & {\em 20.42$\pm$0.13} & {\em 16.66$\pm$0.04} & {\em 15.01$\pm$0.01} & 19.69$\pm$0.04 & 18.45$\pm$0.04     & 
17.26$\pm$0.03  \\
   -   &  016  & 266.56358 & -28.82955 & 17.08$\pm$0.05 & 13.77$\pm$0.05 & 12.26$\pm$0.02 & 16.00$\pm$0.01 & 15.02$\pm$0.02     & 
14.09$\pm$0.01  \\ 
   -   &  017  & 266.56541 & -28.82941 & 18.28$\pm$0.06 & 14.71$\pm$0.04 & 13.11$\pm$0.02 & 17.27$\pm$0.02 & 16.21$\pm$0.02     & 
15.15$\pm$0.02   \\
   -   &   018 & 266.56497 & -28.82931 & 19.74$\pm$0.08 & 15.99$\pm$0.03 & 14.38$\pm$0.01 & 18.55$\pm$0.02 & 17.44$\pm$0.02     & 
16.38$\pm$0.02  \\
   -   &   024 & 266.56709 & -28.82911 & 17.86$\pm$0.09 & 14.57$\pm$0.05 & 13.11$\pm$0.02 & 16.72$\pm$0.02 & 15.77$\pm$0.02     & 
14.84$\pm$0.02  \\ 
   -   &   026 & 266.56287 & -28.82899 & 18.17$\pm$0.09 & 14.58$\pm$0.05 & 13.13$\pm$0.03 & 17.05$\pm$0.02 & 16.00$\pm$0.02     & 
15.02$\pm$0.02  \\
   -   &   028 & 266.56372 & -28.82879 & 17.69$\pm$0.09 & 14.19$\pm$0.05 & 12.69$\pm$0.02 & 16.58$\pm$0.01 & 15.58$\pm$0.02     & 
14.57$\pm$0.01  \\   
   -   &   029 & 266.56139 & -28.82862 & 15.20$\pm$0.08 & 11.80$\pm$0.05 & 10.29$\pm$0.02 & 14.16$\pm$0.01 & 13.13$\pm$0.01     & 
12.13$\pm$0.01  \\ 
   -   &   031 & 266.56325 & -28.82857 & 17.22$\pm$0.07 & 13.88$\pm$0.06 & 12.38$\pm$0.02 & 16.24$\pm$0.01 & 15.23$\pm$0.02     & 
14.25$\pm$0.01  \\  
   -   &   032 & 266.56574 & -28.82856 & 18.57$\pm$0.07 & 15.21$\pm$0.04 & 13.68$\pm$0.02 & 17.48$\pm$0.02 & 16.49$\pm$0.02     & 
15.55$\pm$0.02  \\
   -   &   033 & 266.56921 & -28.82847 & 17.74$\pm$0.08 & 14.41$\pm$0.04 & 13.09$\pm$0.02 & 16.65$\pm$0.01 & 15.68$\pm$0.02     & 
14.75$\pm$0.01  \\ 
   -   &   035 & 266.56223 & -28.82820 & 18.69$\pm$0.08 & 15.20$\pm$0.05 & 13.72$\pm$0.02 & 17.73$\pm$0.02 & 16.67$\pm$0.02     & 
15.65$\pm$0.02  \\  
   -   &   037 & 266.55973 & -28.82814 & 20.55$\pm$0.07 & 16.60$\pm$0.05 & 14.99$\pm$0.02 & 19.50$\pm$0.03 & 18.30$\pm$0.03     & 
17.14$\pm$0.03  \\ 
   -   &   039 & 266.56021 & -28.82797 & 17.67$\pm$0.05 & 14.18$\pm$0.04 & 12.64$\pm$0.01 & 16.60$\pm$0.01 & 15.58$\pm$0.02     & 
14.59$\pm$0.01  \\
   -   &   041 & 266.56601 & -28.82800 & 17.89$\pm$0.07 & 14.49$\pm$0.06 & 12.96$\pm$0.02 & 16.83$\pm$0.02 & 15.81$\pm$0.02     & 
14.83$\pm$0.02  \\
   -   &   044 & 266.56231 & -28.82785 & 16.89$\pm$0.05 & 13.24$\pm$0.14 & 11.98$\pm$0.02 & 16.00$\pm$0.01 & 14.73$\pm$0.02     & 
13.95$\pm$0.01  \\
   -   &   045 & 266.55875 & -28.82789 & 21.85$\pm$0.09 & 17.59$\pm$0.05 & 15.61$\pm$0.06 & 20.83$\pm$0.06 & 19.53$\pm$0.06     & 
18.25$\pm$0.05  \\
   -   &   046 & 266.56347 & -28.82783 & 15.90$\pm$0.06 & 12.61$\pm$0.04 & 11.18$\pm$0.02 & 14.88$\pm$0.01 & 13.87$\pm$0.01     & 
12.96$\pm$0.01  \\
   -   &   048 & 266.56281 & -28.82768 & 17.89$\pm$0.07 & 14.43$\pm$0.04 & 12.96$\pm$0.02 & 16.70$\pm$0.02 & 15.72$\pm$0.02     & 
14.75$\pm$0.02  \\
   -   &   050 & 266.56788 & -28.82760 & 17.72$\pm$0.06 & 14.42$\pm$0.05 & 13.00$\pm$0.01 & 16.67$\pm$0.01 & 15.71$\pm$0.02     & 
14.78$\pm$0.01  \\
   -   &   051 & 266.56575 & -28.82755 & 17.02$\pm$0.12 & 13.47$\pm$0.12 & 12.17$\pm$0.08 & 16.23$\pm$0.01 & 15.15$\pm$0.02     & 
14.17$\pm$0.01  \\ 
   -   &   054 & 266.56035 & -28.82745 & 17.31$\pm$0.07 & 13.73$\pm$0.06 & 12.07$\pm$0.01 & 16.19$\pm$0.01 & 15.13$\pm$0.02     & 
14.10$\pm$0.01  \\
   -   &   055 & 266.56547 & -28.82747 & 17.65$\pm$0.10 & 14.13$\pm$0.04 & 12.57$\pm$0.03 & 16.49$\pm$0.01 & 15.47$\pm$0.02     & 
14.47$\pm$0.01  \\ 
   -   &   056 & 266.56258 & -28.82740 & 18.16$\pm$0.07 & 14.72$\pm$0.04 & 13.17$\pm$0.35 & 16.97$\pm$0.02 & 15.95$\pm$0.02     & 
14.95$\pm$0.02   \\
   -   &   058 & 266.56342 & -28.82730 & 18.68$\pm$0.09 & 15.12$\pm$0.04 & 13.61$\pm$0.01 & 17.50$\pm$0.02 & 16.46$\pm$0.02     & 
15.47$\pm$0.02  \\ 
   -   &   059 & 266.56417 & -28.82728 & 18.71$\pm$0.07 & 15.42$\pm$0.03 & 13.97$\pm$0.02 & 17.63$\pm$0.02 & 16.68$\pm$0.02     & 
15.75$\pm$0.02  \\
   -   &   060 & 266.55915 & -28.82728 & 22.48$\pm$0.21 & 17.82$\pm$0.06 & 15.69$\pm$0.06 & 21.88$\pm$0.12 & 20.31$\pm$0.11     & 
18.84$\pm$0.08  \\   
   -   &   061 & 266.55987 & -28.82718 & 21.47$\pm$0.12 & 17.27$\pm$0.04 & 15.58$\pm$0.02 & 20.31$\pm$0.06 & 19.00$\pm$0.06     & 
17.76$\pm$0.04  \\ 
   -   &   062 & 266.56912 & -28.82720 & 19.33$\pm$0.08 & 16.07$\pm$0.03 & 14.70$\pm$0.02 & 18.28$\pm$0.02 & 17.34$\pm$0.02     & 
16.45$\pm$0.02  \\  
   -   &   064 & 266.56503 & -28.82712 & 18.43$\pm$0.09 & 15.09$\pm$0.04 & 13.52$\pm$0.03 & 17.36$\pm$0.02 & 16.36$\pm$0.02     & 
15.38$\pm$0.01  \\
   -   &   065 & 266.56232 & -28.82712 & 18.88$\pm$0.05 & 15.36$\pm$0.04 & 13.81$\pm$0.01 & 17.69$\pm$0.02 & 16.65$\pm$0.02     & 
15.64$\pm$0.02  \\
   -   &   069 & 266.55962 & -28.82698 & 17.59$\pm$0.08 & 13.80$\pm$0.06 & 12.07$\pm$0.02 & 16.38$\pm$0.01 & 15.26$\pm$0.02     & 
14.19$\pm$0.01  \\
   -   &   072 & 266.56381 & -28.82683 & {\em 17.58$\pm$0.19} & {\em 14.12$\pm$0.12} & {\em 12.92$\pm$0.02} & 16.52$\pm$0.01 & 15.55$\pm$0.02     & 
14.56$\pm$0.01  \\
   -   &   073 & 266.56225 & -28.82683 &    -       &    -       & {\em 12.34$\pm$0.04} & 16.43$\pm$0.01 & 15.37$\pm$0.02     & 14.32$\pm$0.01  
\\ 
   -   &   074 & 266.56050 & -28.82679 & 16.67$\pm$0.07 & 12.98$\pm$0.05 & 11.35$\pm$0.02 & 15.65$\pm$0.01 & 14.55$\pm$0.02     & 
13.40$\pm$0.01  \\
   -   &   076 & 266.55893 & -28.82645 & 17.25$\pm$0.07 & 13.20$\pm$0.05 & 10.88$\pm$0.01 & 16.14$\pm$0.01 & 14.81$\pm$0.02     & 
13.83$\pm$0.01  \\   
   -   &   081 & 266.56660 & -28.82617 & 19.14$\pm$0.07 & 15.41$\pm$0.04 & 13.90$\pm$0.02 & 17.88$\pm$0.02 & 16.83$\pm$0.02     & 
15.81$\pm$0.02  \\      
   -   &   088 & 266.55964 & -28.82601 & 18.99$\pm$0.09 & 15.42$\pm$0.05 & 13.88$\pm$0.02 & 17.86$\pm$0.01 & 16.83$\pm$0.02     & 
15.82$\pm$0.02  \\ 
   -   &   089 & 266.56262 & -28.82588 & 16.70$\pm$0.05 & 13.21$\pm$0.04 & 11.59$\pm$0.01 & 15.84$\pm$0.01 & 14.78$\pm$0.02     & 
13.72$\pm$0.01  \\ 
   -   &   090 & 266.56210 & -28.82590 & 15.96$\pm$0.08 & 12.45$\pm$0.05 & 10.82$\pm$0.02 & 14.86$\pm$0.01 & 13.85$\pm$0.01     & 
12.84$\pm$0.01  \\
   -   &   092 & 266.56958 & -28.82580 & 19.35$\pm$0.06 & 16.03$\pm$0.04 & 14.59$\pm$0.01 & 18.18$\pm$0.02 & 17.23$\pm$0.02     & 
16.32$\pm$0.02  \\   
   -   &   094 & 266.56563 & -28.82573 & 19.06$\pm$0.09 & 15.42$\pm$0.05 & 13.95$\pm$0.03 & 17.86$\pm$0.02 & 16.83$\pm$0.02     & 
15.82$\pm$0.02  \\ 
   -   &   097 & 266.56960 & -28.82561 & 19.62$\pm$0.04 & 16.32$\pm$0.04 & 14.87$\pm$0.01 & 18.59$\pm$0.02 & 17.61$\pm$0.02     & 
16.68$\pm$0.02  \\
   -   &   100 & 266.56323 & -28.82538 & 14.98$\pm$0.10 & 11.43$\pm$0.05 &  9.91$\pm$0.02 & 14.14$\pm$0.01 & 13.05$\pm$0.01     & 
12.01$\pm$0.01  \\ 
   -   &   101 & 266.56276 & -28.82538 & 18.33$\pm$0.08 & 14.76$\pm$0.03 & 13.12$\pm$0.03 & 17.30$\pm$0.02 & 16.23$\pm$0.02     & 
15.17$\pm$0.02  \\ 
   -   &   103 & 266.55910 & -28.82539 & 18.28$\pm$0.07 & 14.84$\pm$0.05 & 13.28$\pm$0.02 & 17.16$\pm$0.02 & 16.16$\pm$0.02     & 
15.19$\pm$0.02  \\ 
   -   &   104 & 266.56808 & -28.82536 & 19.29$\pm$0.05 & 15.80$\pm$0.05 & 14.34$\pm$0.03 & 18.24$\pm$0.02 & 17.21$\pm$0.02     & 
16.21$\pm$0.02  \\
   -   &   105 & 266.56501 & -28.82529 & 18.49$\pm$0.07 & 15.01$\pm$0.05 & 13.57$\pm$0.03 & 17.37$\pm$0.02 & 16.35$\pm$0.02     & 
15.35$\pm$0.02  \\
   -   &   106 & 266.56797 & -28.82513 & 17.91$\pm$0.06 & 14.55$\pm$0.05 & 13.06$\pm$0.02 & 16.93$\pm$0.02 & 15.92$\pm$0.02     & 
14.92$\pm$0.02  \\
   -   &   107 & 266.56697 & -28.82505 & 19.68$\pm$0.06 & 15.97$\pm$0.05 & 14.35$\pm$0.01 & 18.56$\pm$0.02 & 17.43$\pm$0.02     & 
16.37$\pm$0.02  \\
   -   &   109 & 266.56243 & -28.82487 & 18.30$\pm$0.10 & 14.86$\pm$0.04 & 13.32$\pm$0.02 & 17.15$\pm$0.02 & 16.17$\pm$0.02     & 
15.20$\pm$0.02  \\
   -   &   111 & 266.56241 & -28.82468 & 17.60$\pm$0.12 & 14.07$\pm$0.04 & 12.46$\pm$0.03 & 16.40$\pm$0.01 & 15.39$\pm$0.02     & 
14.38$\pm$0.02  \\
   -   &   118 & 266.56692 & -28.82426 & 17.27$\pm$0.06 & 13.56$\pm$0.05 & 11.91$\pm$0.01 & 16.29$\pm$0.01 & 15.12$\pm$0.02     & 
13.97$\pm$0.01  \\
   -   &   120 & 266.56913 & -28.82426 & 20.01$\pm$0.08 & 16.39$\pm$0.04 & 14.76$\pm$0.01 & 18.92$\pm$0.02 & 17.83$\pm$0.02     & 
16.77$\pm$0.02  \\ 
   -   &   122 & 266.55933 & -28.82406 & 17.43$\pm$0.06 & 13.92$\pm$0.06 & 12.33$\pm$0.02 & 16.34$\pm$0.01 & 15.28$\pm$0.02     & 
14.27$\pm$0.01  \\
   -   &   125 & 266.56038 & -28.82397 & 19.67$\pm$0.10 & 15.95$\pm$0.11 & 14.51$\pm$0.11 & 18.59$\pm$0.02 & 17.56$\pm$0.02     & 
16.57$\pm$0.02  \\
   -   &   126 & 266.56899 & -28.82400 & 18.28$\pm$0.06 & 14.67$\pm$0.04 & 13.02$\pm$0.01 & 17.22$\pm$0.02 & 16.14$\pm$0.02     & 
15.05$\pm$0.02  \\
   -   &   128 & 266.56082 & -28.82383 & 18.66$\pm$0.06 & 15.35$\pm$0.07 & 13.79$\pm$0.03 & 17.64$\pm$0.02 & 16.61$\pm$0.02     & 
15.64$\pm$0.02  \\
   -   &   130 & 266.56847 & -28.82364 & 15.45$\pm$0.06 & 14.83$\pm$0.07 & 14.50$\pm$0.03 & 15.05$\pm$0.01 & 14.93$\pm$0.02     & 
14.74$\pm$0.01  \\
   -   &   132 & 266.56011 & -28.82348 & 18.75$\pm$0.06 & 15.15$\pm$0.03 & 13.53$\pm$0.02 & 17.77$\pm$0.01 & 16.68$\pm$0.02     & 
15.59$\pm$0.02  \\
   -   &   134 & 266.56366 & -28.82345 & 18.82$\pm$0.07 & 15.13$\pm$0.05 & 13.58$\pm$0.01 & 17.66$\pm$0.02 & 16.61$\pm$0.02     & 
15.58$\pm$0.02  \\
   -   &   139 & 266.56919 & -28.82307 & 21.22$\pm$0.11 & 15.99$\pm$0.04 & 13.73$\pm$0.03 & 18.51$\pm$0.02 & 17.36$\pm$0.02     & 
16.25$\pm$0.02  \\
   -   &   141 & 266.56245 & -28.82290 & 17.59$\pm$0.05 & 14.08$\pm$0.05 & 12.54$\pm$0.03 & 16.48$\pm$0.01 & 15.47$\pm$0.02     & 
14.44$\pm$0.01  \\
   -   &   143 & 266.56675 & -28.82261 & 16.25$\pm$0.10 & 12.56$\pm$0.05 & 10.86$\pm$0.01 & 15.07$\pm$0.01 & 13.99$\pm$0.01     & 
12.94$\pm$0.01  \\
   -   &   144 & 266.56388 & -28.82237 & 17.54$\pm$0.08 & 14.10$\pm$0.05 & 12.62$\pm$0.02 & 16.44$\pm$0.01 & 15.45$\pm$0.02     & 
14.49$\pm$0.01  \\
   -   &   145 & 266.56671 & -28.82239 & 18.36$\pm$0.07 & 14.79$\pm$0.03 & 13.18$\pm$0.03 & 17.34$\pm$0.02 & 16.29$\pm$0.02     & 
15.25$\pm$0.02  \\
   -   &   148 & 266.56274 & -28.82188 & 17.35$\pm$0.06 & 13.83$\pm$0.05 & 12.25$\pm$0.02 & 16.25$\pm$0.01 & 15.15$\pm$0.02     & 
14.17$\pm$0.01  \\ 
   -   &   149 & 266.55838 & -28.82178 & 17.54$\pm$0.24 & 13.61$\pm$0.17 & 12.36$\pm$0.11 & 16.62$\pm$0.01 & 15.50$\pm$0.02     & 
14.49$\pm$0.01  \\
   -   &   152 & 266.56696 & -28.82158 & 20.34$\pm$0.08 & 16.36$\pm$0.05 & 14.51$\pm$0.02 & 19.16$\pm$0.02 & 17.92$\pm$0.02     & 
16.75$\pm$0.02  \\
   -   &   154 & 266.56184 & -28.82148 & 18.65$\pm$0.08 & 14.88$\pm$0.05 & 13.30$\pm$0.02 & 17.41$\pm$0.01 & 16.32$\pm$0.01     & 
15.27$\pm$0.02  \\
   -   &   157 & 266.56690 & -28.82138 & 21.80$\pm$0.10 & 17.04$\pm$0.04 & 14.90$\pm$0.02 & 21.20$\pm$0.05 & 19.58$\pm$0.05     & 
17.91$\pm$0.03  \\
   -   &   159 & 266.56638 & -28.82125 & 19.94$\pm$0.08 & 16.07$\pm$0.04 & 14.31$\pm$0.02 & 18.83$\pm$0.02 & 17.66$\pm$0.02     & 
16.51$\pm$0.02  \\
       &                   &           &                &                &                &                &                    &       
                 \\
 \multicolumn{2}{c}{WR 102ca}  & 266.55436 & -28.82363 & 17.07$\pm$0.08 & 13.16$\pm$0.07 & 10.90$\pm$0.01 & 16.14$\pm$0.01 & 
14.84$\pm$0.02     & 
13.90$\pm$0.01  \\
\end{longtable}
{Objects for which no photometry is available are qF76 (WC9; this work), qF269 (OB I; Fi99a), qF301 ($<$OB I; Fi99a), qF311 (B1-3 I; 
Fi99a), qF362 (LBV; Geballe et al. \cite{geballe00}), LHO23 (O3-6 I-II; Li09), LHO70 (O7-8 If; Li09) and G0.120-0.048 (LBV; Mauerhan 
et al. \cite{mauerhan10a}). qF353E, -381 and -406 were outside the NICMOS field-of-view and hence lack F110W, F160W and F205W 
magnitudes. The following sources had multiple  counterparts within  2.5pixels of their position in the NICMOS data; qF276 (two 
sources  with one 2.5mag fainter; brighter selected), LHO15 (two sources with a difference of 0.3m; closest selected), LHO64 (two 
sources with one 3.4mag fainter; brighter selected), LHO72 (two sources with one 1.2mag fainter; closest selected) and  LHO73 (two 
sources within one pixel with broadly comparable magnitudes; closest selected). Caution should therefore be applied regarding the 
F110W, F160W and F205W magnitudes for LHO15, -72 and -73; these are given in italics. Finally no NICMOS counterpart to LHO139 was found within 
9 pixels of the anticipated position.
}}
  
\begin{figure*}
\includegraphics[width=12cm,angle=0]{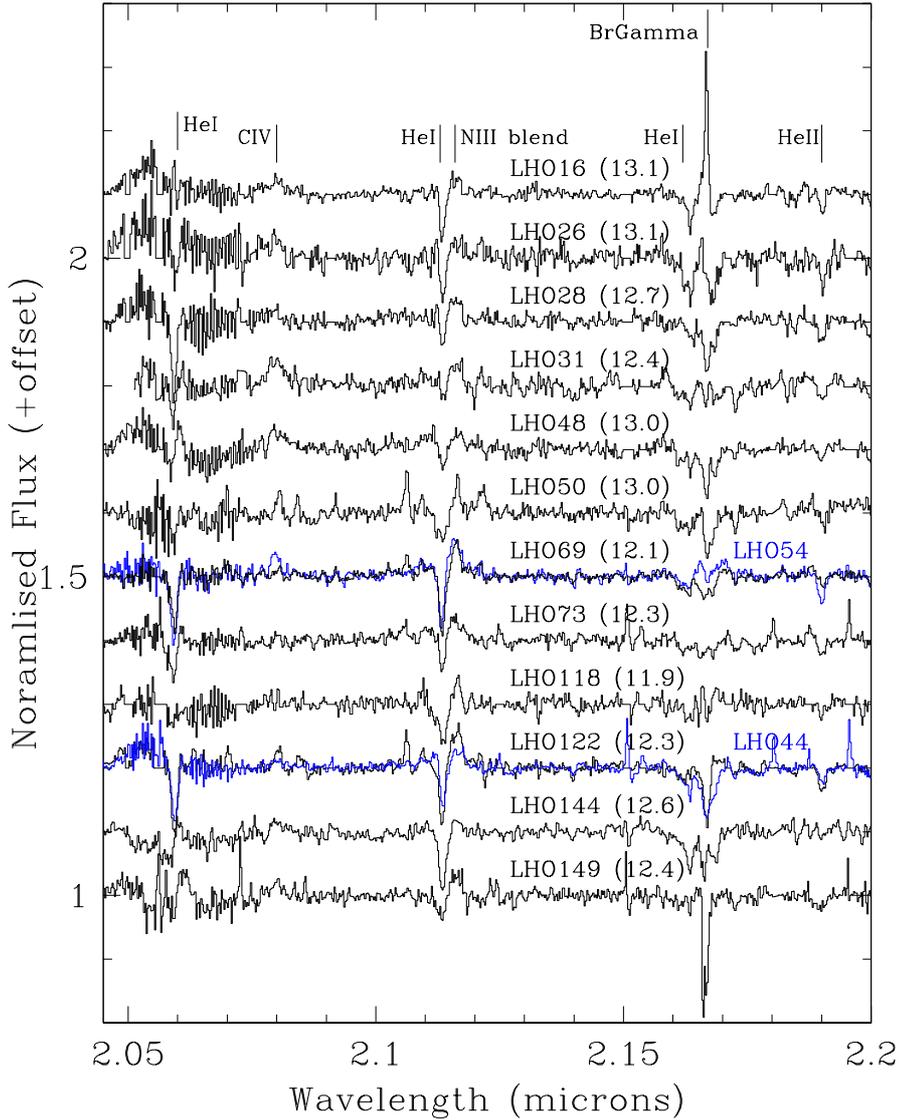}
\caption{Additional cluster supergiants (black) of spectral type O7-8  by virtue of
 He\,{\sc ii} 2.189$\mu$m absorption and/or C\,{\sc iv} 2.079$\mu$m emission. Spectra of cluster members of comparable spectral type overplotted in blue for comparison. The weakness of He\,{\sc ii} 2.189$\mu$m in LHO118, -144 and -149 marks them as of the latest spectral type, with C\,{\sc iv} 2.079$\mu$m emission also apparently absent in LHO118. Note apparent infilling of Br$\gamma$ in LHO31, -69, -73, -118 and -144. Emission in the Br$\gamma$ profile of LHO16 is suspected to be part nebular in origin, while the depth of Br$\gamma$ absorption in LHO149 is probably due to subtraction of the nebular-contaminated sky frame.  HST/NICMOS F205W magnitudes are given in parentheses (Table A.1). }
\end{figure*}

\begin{figure*}
\includegraphics[width=12cm,angle=0]{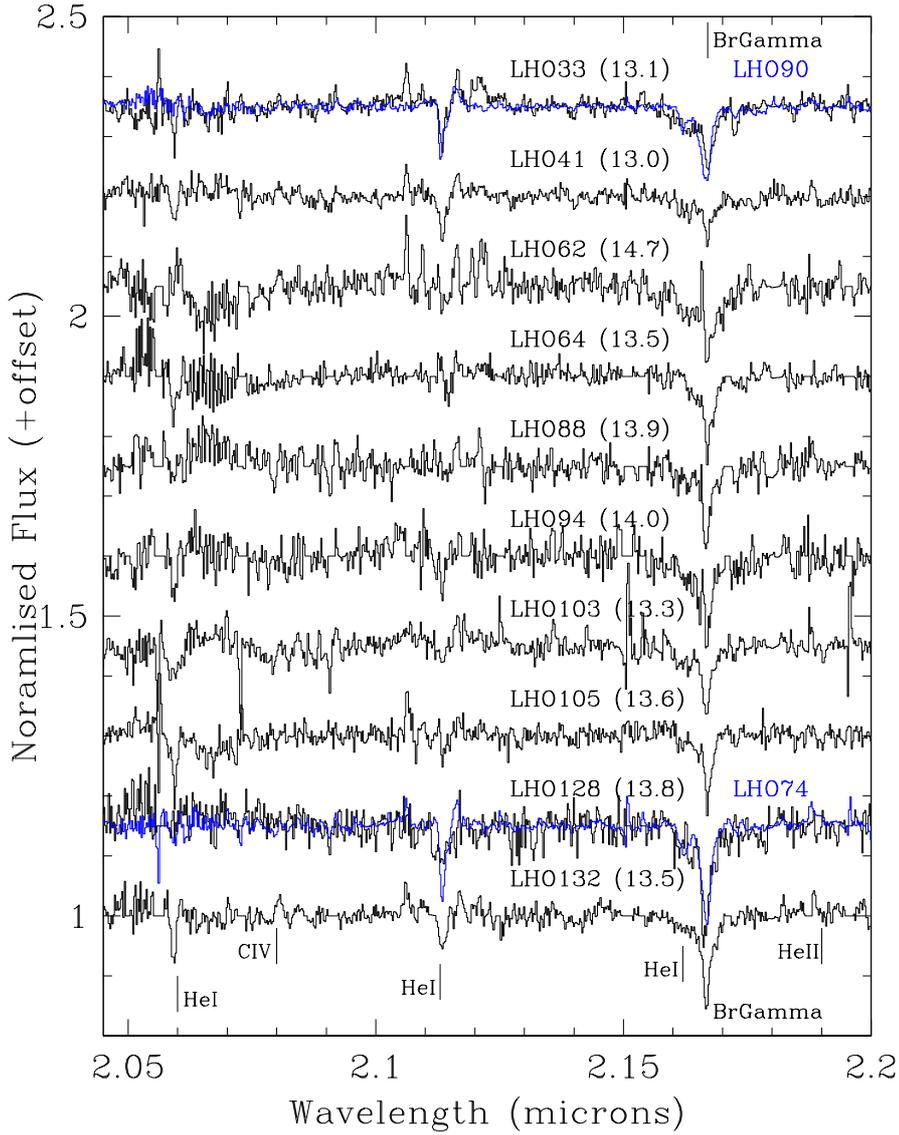}
\caption{Additional cluster supergiants (black) which lack He\,{\sc ii} 2.189$\mu$m photospheric absorption. 
Overplotted in blue are the spectra of two cluster O9-B0Ia stars suggesting comparable classifications for LHO33, -41, -128 and -132. The lower S/N of the remaining objects results in a generic OB star classification.
HST/NICMOS F205W magnitudes are given in parentheses (Table A.1). }
\end{figure*}

\end{document}